\documentclass[11pt,number,sort&compress]{amsart}

\usepackage{amssymb}
\usepackage{amsthm}

\usepackage{lineno}

\usepackage[utf8]{inputenc}
\usepackage{amsmath}
\usepackage{array, amsfonts}

\usepackage{enumerate}
\usepackage[bookmarks=true]{hyperref}
\usepackage{enumitem}
\usepackage{mathrsfs}
\usepackage{stackengine}
\usepackage{bookmark}
\usepackage{mathtools}
\usepackage{todonotes}
\usepackage{graphicx}
\usepackage{multicol}
\usepackage{caption}
\usepackage{subcaption}

\usepackage{fancyhdr}

\usepackage{bm}

\usepackage[foot]{amsaddr}

\usepackage{color}
\definecolor{darkgreen}{rgb}{0,0.5,0}

\theoremstyle{definition}
\newtheorem{definition}{Definition}[section]
\newtheorem{notation}[definition]{Notation}

\newtheorem*{CCP}{Complexity caging principle/hypothesis}

\theoremstyle{theorem}
\newtheorem{lemma}{Lemma}[section]
\newtheorem{theorem}[lemma]{Theorem}

\fancypagestyle{plain}{
	\fancyhead{}
	\fancyhead[C]{ \textbf{Extended version} }
}
\begin{document}

\title{ A Simplicity Bubble Problem\\ in Formal-Theoretic Learning Systems %\tnoteref{t1} 
}

\author{Felipe S. Abrah\~{a}o}
\address[Felipe S. Abrah\~{a}o, Itala M. L. D'Ottaviano]{Centre for Logic, Epistemology and the History of Science (CLE), University of Campinas (UNICAMP), Brazil.}
\email{fabrahao@unicamp.br}
\address[Felipe S. Abrah\~{a}o, Fabio Porto, Klaus Wehmuth]{National Laboratory for Scientific Computing (LNCC) -- Petropolis, RJ -- Brazil }

\address[Felipe S. Abrah\~{a}o, Hector Zenil]{Algorithmic Nature Group, Laboratoire de Recherche Scientifique (LABORES) for the Natural and Digital Sciences, Paris, France.}

\author{Hector Zenil}  
\address[Hector Zenil]{
	Machine Learning Group, Department of Chemical Engineering and Biotechnology, The University of Cambridge, U.K..\newline
	Oxford Immune Algorithmics, England, U.K..\newline
	Algorithmic Dynamics Lab, Unit of Computational Medicine, Department of Medicine Solna, Center for Molecular Medicine, Karolinska Institute, Stockholm, Sweden.
}
\email{hector.zenil@cs.ox.ac.uk}

\author[Fabio Porto]{Fabio Porto} 
\email{fporto@lncc.br}

\author{Michael Winter}
%\address[Michael Winter]{Max Planck Institute for the History of Science, Germany}
\email{mwinter@unboundedpress.org}

\author[Klaus Wehmuth]{Klaus Wehmuth} 
\email{klaus@lncc.br}

\author{Itala M. L. D'Ottaviano}
\email{itala@unicamp.br}

\maketitle\thispagestyle{plain}

\begin{abstract}
When mining large datasets in order to predict new data, limitations of the principles behind statistical machine learning pose a serious challenge not only to the Big Data deluge, but also to the traditional assumptions that data generating processes are biased toward low algorithmic complexity. 
Even when one assumes an underlying algorithmic-informational bias toward simplicity in finite dataset generators, we show that  
current approaches to machine learning (including deep learning, or any formal-theoretic hybrid mix of top-down AI and statistical machine learning approaches), can always be deceived, naturally or artificially, by sufficiently large datasets. 
In particular, we demonstrate that, for every learning algorithm (with or without access to a formal theory), 
there is a sufficiently large dataset size above which the algorithmic probability of an unpredictable deceiver is an upper bound (up to a multiplicative constant that only depends on the learning algorithm) for the algorithmic probability of any other larger dataset.
In other words, very large and complex datasets can deceive learning algorithms into a ``simplicity bubble'' as likely as any other particular non-deceiving dataset.
These deceiving datasets guarantee that any prediction effected by the learning algorithm will unpredictably diverge from the high-algorithmic-complexity globally optimal solution while converging toward the low-algorithmic-complexity locally optimal solution, although the latter is deemed a global one by the learning algorithm.
We discuss the framework and additional empirical conditions to be met in order to circumvent this deceptive phenomenon, moving away from statistical machine learning towards a stronger type of machine learning based on, and motivated by, the intrinsic power of algorithmic information theory and computability theory.
\end{abstract}

%
%%%%Graphical abstract
%%%\includegraphics{grabs}
%%\item Research highlight 1
%%\item Research highlight 2
%\begin{keyword}
%%% keywords here, in the form: keyword \sep keyword
%algorithmic information \sep computability theory \sep machine learning \sep Big Data \sep bias toward simplicity \sep data analysis
%
%%% PACS codes here, in the form: \PACS code \sep code
%
%%% MSC codes here, in the form: \MSC code \sep code
%%% or \MSC[2008] code \sep code (2000 is the default)
%
%\end{keyword}
%
%\end{frontmatter}

%\pagebreak

%\linenumbers

\pagebreak
\section{Introduction}
\label{labelIntro}

The success of statistical machine learning is undeniable, but the expectation of what it can achieve requires an investigation into its fundamental limitations. %are based upon for finding globally optimal parameters that minimise the generalisation (or prediction) errors.
One problem often encountered in computational analyses of large datasets is that data mining algorithms are prone to finding statistically significant correlations that are in fact spurious (i.e., correlations that would have occurred in completely randomly generated data anyway) \cite{Calude2017}. Data scientists attempt to mitigate this consequence by empirically defining features to extract, but generally, blind or naive applications of data mining algorithms to sufficiently large datasets increase the chance of these algorithms finding spurious regularities.
This gives rise to a ``paradox'' that underlies the Big Data paradigm \cite{Smith2020}: although computational analysis benefits from larger datasets that are accurate and fine grained, the larger the dataset the more likely it is that spurious correlations will be discovered.
%, with the consequence that the data mining process yields meaningless conclusions.
In other words, ``too much information tends to behave like very little information'' \cite{Calude2017}.

In machine learning, a central problem is to avoid underfitting and overfitting: that is, when the value of the loss/cost/error function obtained from the training set is considered to be high; or when the error on the test set is considered to be relatively higher than the training error, respectively. Underfitting can be solved by a number of simple methods such as increasing the model capacity or complexity. However, solving overfitting demands much more sophisticated methods that succumb to Parsimony, Occam's razor, or a bias toward simplicity. This consequence is considered by some as a fundamental principle or heuristic in methods that try to solve the problem of overfitting \cite{Goodfellow2016,Witten2017}.
Indeed, recent methods and results that consider this consequence as fundamental have shown predictive capability for both traditional learning methods in differential spaces (by loss functions, optimisation, and regularisation that directly stem from the bias toward low algorithmic complexity), but also in non-differential spaces that generally perform poorly using (or, in some cases, cannot in principle be tackled by) traditional methods \cite{HernandezOrozco2021}.
Due to the inherent bias toward simplicity in the universal probability distribution of computably generated objects \cite{Calude2002}, models and parameters that yield a perfect fit with a minimum training error for every data point in the training set tend to have greater algorithmic complexity.
Thus, models and parameters with low algorithmic probability would naturally be penalised in the parameter and hyperparameter optimisation stage of the learning process, and such penalisation would in turn avoid overfitting.

In this article we investigate fundamental limitations and conditions in a context that combines both the ``too much information tends to behave like very little information'' phenomenon and the bias toward simplicity.
In particular, given an arbitrarily chosen machine learning algorithm, we study worst-case scenarios in which overfitting is expected to always occur once the datasets get sufficiently large, even if the datasets are generated following the universal distribution.
This reveals a fundamental limitation in the Big Data paradigm that not only corroborates the results in \cite{Calude2017} and \cite{HernandezOrozco2021}, but also demonstrates the necessity of new algorithmic information-based conditions to avoid finding ``spurious optimal models'' when mining very large datasets.

The problem of learning algorithms' solutions getting stuck in local optima is a well-known problem to be faced both in practice and in theory.
In this regard, our results not only demonstrate conditions under which such a problem is guaranteed to occur, but also that such a limitation is always expected to occur when applying current learning methods, should they only rely on statistics to deal with large enough datasets of arbitrarily high complexity.  
To overcome this, one necessarily needs to guarantee that additional properties are met; and to develop new learning algorithms that stem from constraining the scope of the algorithmic information content in the datasets to be investigated (see Section~\ref{sectionDiscussion}).

Our primary results are two proofs (Theorem~\ref{thmDeceivingprobability} which follows from Theorem~\ref{thmDeceiverDatasets}) showing that, if a locally optimal model found by an arbitrarily chosen machine learning algorithm using a sufficiently large dataset has an algorithmic complexity arbitrarily smaller than that of the available data, the algorithmic probability of the deceiving dataset still dominates the algorithmic probability of any other randomly generated dataset of size greater than or equal to the size of the deceiving dataset.
On the one hand, this enables the deceiving dataset to limit the learning algorithm's capabilities of accurately finding the globally optimal model.
On the other hand, the locally optimal model can be found with arbitrary precision by the learning algorithm.
As formalized in Section~\ref{sectionRandomnessandmachinelearning}, a \emph{deceiving dataset} is one for which the learning algorithm (in addition to any formal theory and performance measure one may choose to embed into this algorithm) generates an optimal model that completely satisfies the assumed performance criteria to be considered a global one, but that in fact is a locally optimal model. 
%Thus, in this scenario, overfitting is guaranteed to occur.

Formally,
we demonstrate in Theorem~\ref{thmDeceivingprobability} that, for every learning algorithm $ \mathrm{P} $, there are sufficiently large datasets $ \mathrm{D}_{total} = \mathrm{D}_a \cup \mathrm{D}_{new} $ whose algorithmic probability of the respective $ \mathrm{D}_a $ being an unpredictable deceiver (which satisfies Theorem~\ref{thmDeceiverDatasets} with $ \left| \mathrm{D}_a \right| \geq k $) is higher than (except for a multiplicative constant that depends on $ \mathrm{P} $) the algorithmic probability of any other particular dataset of size $ \geq k $.
$ \mathrm{P} $ is the arbitrary learning algorithm deceived by the available dataset $ \mathrm{D}_a $.
The available dataset $ \mathrm{D}_a $ (from which $ \mathrm{P} $ calculates an optimal model $ \mathrm{M}_{ \left( \mathrm{P} , \mathrm{D}_a \right) } $) is sufficiently large so that there is $ k \in \mathbb{N} $ such that 
$ \left| \mathrm{D}_a \right| 
\geq  
k 
= 
BB\left(
\mathbf{K}\left( \mathrm{D}'_{total} \right) 
- 
\mathbf{O}\left( \mathbf{K}\left( \mathrm{P} \right)
\right) \right) $, where $ \mathrm{D}'_{total} $ is any dataset that satisfies Theorem~\ref{thmDeceiverDatasets},
$ BB\left( \cdot \right) $ is a Busy Beaver function,
and 
$ \mathbf{K}\left( \cdot \right) $ is the (prefix) algorithmic complexity.
%and
%$ \delta > 0  $ is a rational number such that $ \delta $ is smaller than or equal to the maximum generalisation/prediction error with respect to $ \mathrm{M}_{ \left( \mathrm{P} , \mathrm{D}'_a \right) } $, for any $ \mathrm{D}'_a $.

The theorems are proven by showing that for any given learning algorithm, one can always construct a sufficiently large dataset (which includes the training set, test set, validation set, etc.), such that the prediction error (i.e., the overall generalisation error taking into account fresh data) cannot be minimised beyond the training error from already available data.
The key step to achieving this limitation is to ensure that data is generated in such a way that the joint algorithmic complexity of the learning algorithm combined with the available dataset, and any locally optimal model is smaller than the algorithmic complexity of the globally optimal model by a large enough constant.

This particular way in which large enough datasets can deceive learning algorithms into accurately finding local optima while still being over-fitted has to do with a phenomenon that appears from a complexity-centered interplay between data and algorithms.
As available data becomes sufficiently large, its underlying generating process (or mechanism) likely becomes more complex than the learning algorithm.
In this scenario, the data generating process may ``trap'' the learning algorithm into a ``simplicity bubble'' in which any solution found from the available data alone is necessarily not complex enough to match the complexity of the actual data generating process. 
Only by adding a larger amount of (unpredictable) fresh data, can the learning algorithm hypothetically be capable of ``bursting'' this bubble. 
We call this phenomenon the \emph{simplicity bubble effect}.

Section~\ref{sectionBackground} sets the background for our forthcoming results and explains the specific issues in the literature that motivate this article.
A brief overview of the main concepts and remarking features of machine learning is presented in Section~\ref{sectionMachinelearning}. In Section~\ref{sectionMachinelearninginlargedatasets},
we discuss fundamental problems related to machine learning and large datasets.
Then in Section~\ref{sectionMachinelearningandalgorithmicinformation}, we present previous methods and results for machine learning based on algorithmic information theory.
Section~\ref{sectionComputabilitytheoryandmachinelearning} introduces a novel formalisation of machine learning in computability theory, and Section~\ref{sectionRandomnessandmachinelearning} formalises the concept of unpredictable deceivers in this context.
Our main results are demonstrated in Section~\ref{sectionResults}.
Finally, in Section~\ref{sectionDiscussion}, we conclude with a discussion of additional assumptions to avoid the learning-deceiving phenomenon of the simplicity bubble effect.

\section{Background}\label{sectionBackground}

\subsection{Basics concepts in machine learning}\label{sectionMachinelearning}

Traditionally, in machine learning a \emph{model} $ \mathrm{M} $ is an equation, function, procedure, or algorithm defined upon 
%Let $ \mathrm{M} $ be an arbitrary \emph{model} defined 
the set of fixed parameters $ \boldsymbol{ \theta } = \left\{ \theta_1, \theta_2 , \dots , \theta_n , \dots \right\} $ and the set of hyperparameters $ \pmb{ \gamma } = \left\{ \gamma_1, \gamma_2 , \dots , \gamma_{n'} \right\} $, where $ n, n' \in \mathbb{N} $ depend on $ \mathrm{M} $, model which executes the task the learning process is trying to learn \cite{Goodfellow2016}.
In this manner, one says a model $ \mathrm{M} $ is \emph{optimal} if its parameters (and hyperparameters) are such that the model better fits the true results the task should return.
When $ \left| \boldsymbol{ \theta } \right| $ is finite and fixed the model is called \emph{parametric}.
Otherwise, it is called \emph{non-parametric} \cite{Goodfellow2016,Bishop2006,Murphy2012}.

For example, in the case of a classification task, one has a set of objects $ X = \left\{ \mathbf{x}_1, \dots , \mathbf{x}_k \right\} $, a set of categories $ C = \left\{ c_1 , \dots , c_m \right\} $, where $ \mathbf{x}_i \in \mathbb{R}^{< \infty} $ and $ c_j \in \mathbb{N} $, and a model defined by the function $ \mathrm{M} \colon X \to C $ \cite{HernandezOrozco2021,Goodfellow2016,Murphy2012}.
One tries to find the best model $ \mathrm{M} $ such that $ \mathrm{M}\left( \mathbf{x}_i \right) = c_j $ holds with the highest probability iff the vector (or object) $ \mathbf{x}_i $ belongs to the category encoded by the value $ c_j $.

In the case of a \emph{regression} task (which the present article is devoted to), $ X $ is a set of input values from which the model $ \mathrm{M} \colon X \subseteq \mathbb{R}^{ < \infty } \to Y \subseteq \mathbb{R}^{ < \infty } $ is a function\footnote{ Note that, instead of real values in $ \mathbb{R} $, one may also define regression for complex values in $ \mathbb{C} $.} that outputs $ \mathrm{M}\left( \mathbf{x} \right) = \hat{ \mathbf{y} } $.
In more simplistic terms, the general idea of optimality here is that $ \hat{ \mathbf{y} } $ is said to be an optimal approximation of $ \mathbf{ y } = f\left( \mathbf{x} \right)  $ iff $ \left\lVert \hat{ \mathbf{y} } - \mathbf{ y } \right\rVert_{ l^2 }^2  < \epsilon \in \mathbb{R} $, where $ \epsilon $ is a \emph{prediction error} considered to be acceptable with respect to the actual function $ f\left( \cdot \right) $ the model is trying to approximate \cite{HernandezOrozco2021,Goodfellow2016,Murphy2012}.
In Section~\ref{sectionComputabilitytheoryandmachinelearning},  Definition~\ref{defOptimality}, we will further investigate the formalisation of such a notion of optimality and performance measure.
It is important to remark that, in this article, we focus on the tasks that are regression (or prediction) problems.
Whilst generalising the forthcoming results to classification problems is straightforward, we leave for future research the generalisation of our results to other tasks.

A first classical example of regression is \emph{linear regression}, where the model is given by linear combination of weights with a vector plus a bias $ b \in \mathbb{R} $:
\begin{equation}
\mathbf{w}^\mathsf{T} \, \mathbf{x} + b = \hat{ y } \in \mathbb{R}
\text{ .}
\end{equation}
One may think of the weights and biases of such a linear regression as the parameters in the parameter space $ \Theta $ that should be optimised.
There may be other components of a model that are empirically difficult to optimise, and as such they can be assigned to be the \emph{hyperparameters} of a model.
For example, in polynomial regression the degree of the polynomial is a hyperparameter.

Another example of regression is logistic regression \cite{Witten2017,Murphy2012}, using a sigmoid function of the form
\begin{equation}
	\sigma\left( \mathbf{x} \right) = \frac{ 1 }{ 1 + e^{ \left( - \mathbf{w}^\mathsf{T} \, \mathbf{x} - b \right) } } = \hat{ y } \in \mathbb{R}
	\text{ ,}
\end{equation}
and it is often adopted in classification problems.
Being widely employed in \emph{artificial neural networks} (ANNs), a rectified linear unit (or ReLU) \cite{Witten2017,Goodfellow2016,Nielsen2015} of the form
\begin{equation}
\textit{rectify}\left( \mathbf{x} \right) = b +
\mathbf{w}^\mathsf{T}  \max \left\{ \mathbf{ 0 } , \mathbf{W}^\mathsf{T} \mathbf{x} + \mathbf{c}  \right\} = \hat{ y } \in \mathbb{R}
\text{ ,}
\end{equation}
where $ \mathbf{W} $ is a matrix and $ \mathbf{c} $ is a vector,
can also play the role of the ``model'' of an individual neuron (i.e., the activation function, which operates on the output of a layer to transform it into an expected output domain value) \cite{Murphy2012,Witten2017,Bishop2006,Mitchell1997}.

In summary, one can say an ANN is a composite model.
Basically, it is defined as a graph-like combination of other node ``models'' (i.e., the neurons, represented by the vertices) such that each neuron executes an activation function using the weight vectors, weight matrices, biases (i.e., parameters), and the outputs given by each incoming edge from the neighbor neurons, respectively \cite{Goodfellow2016,Bishop2006,Nielsen2015}. 
Thus, even in the case of \emph{deep learning} \cite{Goodfellow2016,Witten2017,Murphy2012}, where there are multiple interconnected distinct layers of neurons, such distributed computation schemes of processing nodes can be understood as a (composite) model defined by the network topology and the collection of nodes with their own respective functions, parameters, weights, biases, etc.

The remarking feature of machine learning is the ability of approximating the optimal model from datasets so that this learnt model could be employed to make generalisations about other data.
The data structure of those sets can be of any sort of objects, such as $n$-dimensional vectors, matrices, or graphs, that can be encoded and with which the learner
%\footnote{A learner is a training algorithm such as a convolutional neural network.} 
can operate and then return model candidates.
In general, the underlying process that generates the datasets is unknown, often taken as stochastic under i.i.d. assumptions \cite{Goodfellow2016}---as we will see in Section~\ref{sectionResults}, this article investigates the case in which the data generating process follows the universal distribution \cite{Calude2002}.

In order to avoid obtaining over-fitted models, several techniques of manipulating the available datasets are employed.
To this end, one for example deploys algorithms and randomisations for sampling and retrieving a training set, test sets, and validation sets from the \emph{available dataset} \cite{Goodfellow2016}.
The \emph{training set} is a dataset on which a learner algorithm can calculate approximate values of the parameters of a model;
the \emph{validation set} is a dataset on which an algorithm can optimise the hyperparameters values;
and the \emph{test set} is a dataset from which an algorithm can calculate the generalisation (or prediction) error given by the optimal model candidate obtained after learning from the training set and validation sets \cite{Goodfellow2016,Witten2017,Murphy2012}.

To finally return an optimal model after evaluating these candidates, a learning algorithm needs to estimate how much the parameters and/or hyperparameters values are wrong with respect to (or diverge from) the actual model that better fits (explain or predicts) the data.
This is calculated by an arbitrarily chosen \emph{error function} (or cost function) whose values are returned by some algorithm (or subprocedure) during the learning process from the available dataset \cite{Mitchell1997,Goodfellow2016}.
An error function calculated on the training set is called the \emph{training error} and on the test set the \emph{generalisation error} (or test error) \cite{Goodfellow2016}.
In cases in which the learner needs to actively change the model parameters in order to minimise the error function value, such as the gradient descent in ANNs, one may require the training error function to be differentiable, and in this case this error function is often called a \emph{loss function}. 
One can for example apply the chain rule and backpropagate the partial derivatives of the weights and biases with the purpose of retrieving the negative gradient, which should change the weights and biases toward values closer to the optimal ones  \cite{Nielsen2015,Goodfellow2016}---the global optimum may be not always found, but the gradient descent can at least move the parameters closer to local optima.

Common examples of error (or loss) functions \cite{Goodfellow2016,Murphy2012,Shcherbakov2013,Bishop2006} that one seeks to minimise are the \emph{mean squared error} (MSE)
\begin{equation}
	\text{MSE}\left( \mathbf{ y } \right)  = \mathbf{ E \left[ \mathnormal{ \left\lVert \hat{ \mathbf{ y } }_i - \mathbf{ y }_i \right\rVert_{ l^2 }^2 } \right] }
	\text{ ,}
\end{equation}
where each $ \hat{ \mathbf{ y } }_i $ is calculated from a single instance input data,
and the \emph{cross-entropy}\footnote{ Or, equivalently, the KL divergence.} 
%(equivalently, but not the same, the KL divergence or the log-likelihood)
\begin{equation}\label{equationCrossentropy}
	\mathcal{ H }\left( y \, || \, \hat{ y } \right) = 
	-
	\mathbf{ E_{ \mathnormal{ y } } \left[ \mathnormal{ \log\left( \hat{ y } \right)  } \right] }
	\text{ ,}
\end{equation}
where in this case both $ y $ and $ \hat{ y } $ are probability distributions.
As we will discuss in Section~\ref{sectionMachinelearningandalgorithmicinformation}, this differentiability condition can be relaxed so as to achieve more widely applicable loss functions.

There are other techniques that the learning algorithm employs during the learning process in order to cope with empirical challenges of guaranteeing a minimisation of the generalisation error without increasing too much the training error.
\emph{Regularisation} often does this by adding a penalty to high values of the parameters in the loss function \cite{Goodfellow2016}. (For example, the weight decay method with the regularisation constant\footnote{ Which is considered to be an hyperparameter.} $ \lambda $ in Equation~\ref{equationRegularisedalgorithmiclossfunction}).

\emph{Cross-validation} partitions the training set and test set when the available dataset is considered to be too small \cite{Goodfellow2016}.
It can further be employed to study the \emph{bias-variance trade-off} so that the generalisation error is minimised while the model is still robust (i.e., the variance does not increase too much) to re-samplings of the training set \cite{Goodfellow2016,Bishop2006}.
\emph{Bagging}, \emph{boosting} and \emph{dropouts} can be employed to compare alternative concurrent models or sub-models, re-sampling training sets from the available dataset \cite{Goodfellow2016,Witten2017,Nielsen2015};

Thus, in general machine learning algorithms can resort on many procedures (either computable or stochastic ones) to return the output (i.e., a model) from the data made available by an external source.
Assuming the learning algorithm always halts for any input data given by the external source, the machine learning process can be understood as a well-defined computable sequence (or combination) of \emph{total computable functions} (and, in some cases, calls to stochastically random number generators) that finally returns a model at the end of this process.

Empirically finding that the learning algorithm is successfully approximating the actual model of the underlying data generating process of the external source gives rise to the idea of an algorithm ``learning'' from the experienced data \cite{Goodfellow2016}.
In this manner, the distinctive features of machine learning process that makes it different from an usual computable function are: 
the presence of an external source that generates the available data, and might also generate new data to be predicted or explained; 
and the existence of an error or \emph{performance measure} that evaluates how well the algorithm learnt from the external source.
In summary, this is grasped by the notion that ``a computer program is said to learn from experience $E$ with respect to some class of tasks $T$ and performance measure $P$, if its performance at tasks in $T$, as measured by $P$, improves with experience $E$'' \cite{Mitchell1997,Goodfellow2016}.
In Section~\ref{sectionComputabilitytheoryandmachinelearning}, we will show how this informal notion can be formalised under an overarching computability-theoretic perspective.

\subsection{Large complex datasets}\label{sectionMachinelearninginlargedatasets}

Learning in large complex datasets $ \mathrm{D} $ potentially introduces spurious patterns or may lead to overfitted models. 
As a matter of fact, although large datasets may exhibit a reasonable quantity of samples of the learning domain, inferences are usually interested in a particular region of such a large domain. In this scenario, finding the weights $W$ (paramaters, or hyperparameters) for a learner $ \mathrm{L} $ that fits
the data in $ \mathrm{D} $ involves optimising the \emph{error function} over a much larger pattern space than a given model $ \mathrm{M} $ may be interested in. As a result, model $ \mathrm{M} $ may perform sub-optimally when inferring from a sub-region of the domain space.

Another interesting point concerns the training cost associated with large datasets, such as $ \mathrm{D} $, and the opportunity to find subspaces $ S \subset \mathrm{D} $, considerably smaller than $ \mathrm{D} $, and producing the same error guarantees as the complete large dataset \cite{Baharan2021}. Thus, one may want to find a subspace $S_j$, with $S=\{S_1, S_2, \dots ,S_n\}$, $ j \in \{1, \dots,  n\} $, and $ \mathrm{D} = \cup_{ i = 1 }^{ n } S_i $, such that two machine learning models $ (\mathrm{M}_i, \mathrm{M}_j) $, using the same learner $ \mathrm{L} $, and trained on the complete domain $S$ and on a subspace $S_j$, respectively, will exhibit prediction errors from inferring from an input $x$,  $\mathrm{M}_i(x)$ and $\mathrm{M}_j(x)$, that differ at most by a threshold $\epsilon$. 
A question that naturally arises is how we may compute the subspaces in $S$ from $ \mathrm{D} $. 

In \cite{Pereira2021}, subspace computation occurs in a pre-processing stage, before training and inferencing. A clustering process partitions the domain $ \mathrm{D} $ into subspaces presenting samples sharing similar values, according to a given distance function. Moreover, for each partition one may compute a representative point, for instance, one whose distance function to all points in the partition is the smallest. Each subspace is a better representation of patterns within its data space. Thus models aiming at predictions in a subspace would fit to patterns in the respective subspace, placing the $loss$ function 
relative to
the patterns of interest. Conversely, once a prediction is to be computed on an input $x$, one may look for models whose trained subspace offers a representative point closest to $x$. 
This, in fact, is in line with the no free lunch theorem \cite{Wolpert1997} for machine learning: 
for every learner, there exists a task at which it fails, even though that task can be successfully learned by another learner.
In other words, no learning algorithm is any better than any
other equally likely learning algorithm \cite{Goodfellow2016}.

To tackle this latter problem, assumptions or heuristics based on variants of the minimum description length, parsimony, Occam's razor, or the bias toward simplicity are employed in machine learning methods and data science \cite{Goodfellow2016,Witten2017,Scholkopf2021}.
By assuming skewed distributions in empirical data gathering that favor simpler models (e.g., by assigning penalties into the error functions, as done in regularisation), one could avoid the average of equally likely datasets and/or models that constitutes one of conditions for the no free lunch theorem.
In this regard, ranging over all computably constructible objects, the algorithmic coding theorem \cite{Calude2002,Chaitin2004} sets a foundational result that connects algorithmic complexity and probability (semi-)measures, formalising the bias toward simplicity in algorithmic information theory (AIT):
\begin{equation}\label{equationACT}
\mathbf{K}\left( x \right) = - \log\left( \sum\limits_{ \mathbf{U}\left( p \right) 
	= x } \frac{ 1 }{ 2^{ \left| p \right| } } \right) \pm \mathbf{O}( 1 )
= - \log\left( \mathbf{m}\left( x \right) \right) \pm \mathbf{O}( 1 )
\text{ ,}
\end{equation}
where
$ p $ denotes a program running on the universal prefix Turing machine $ \mathbf{U} $,
$ \left| p \right| $ is the length of a program $ p $,
$ \mathbf{m}\left( \cdot \right) $ is a maximal computably enumerable semimeasure, 
and $ \sum\limits_{ \mathbf{U}\left( p \right) 
	= x } 2^{ - \left| p \right| }  $ 
(called \emph{universal a priori probability} of $ x $)
indicates the probability of obtaining $ x $ from an i.i.d. random generation of prefix-free (or self-delimited) programs.
Indeed, as we will discuss in the next Section~\ref{sectionMachinelearningandalgorithmicinformation}, recent methods and results have shown the contributions of this formal version of bias toward simplicity.

\subsection{Algorithmic information and model discovery}\label{sectionMachinelearningandalgorithmicinformation}

A suite of algorithms based on such first mathematical principles (able to characterise the concepts of information and randomness) were shown to enable the study of causal inference, reprogrammability, and complexity of evolving systems~\cite{Zenil2018,Zenil2019}. 
In~\cite{HernandezOrozco2021}, it was shown how those tools can be introduced and exploited in the context of machine learning to help bring together and expand apparently disparate areas of current AI research. For example, it was shown that these model-driven approaches require less training data and can be e.g. more generalisable as they show greater resilience to random attacks. 
Hernández-Orozco et al.~\cite{HernandezOrozco2021} investigated the shape of a discrete
algorithmic space when performing regression or classification using a \emph{loss function} parametrised by algorithmic complexity, as e.g. in the regularised loss function \cite{HernandezOrozco2021} for non-differentiable spaces
\begin{equation}\label{equationRegularisedalgorithmiclossfunction}
	J_{ K }\left( X \times Y , \mathrm{M} , \lambda \right) = 
	\lambda \mathbf{K}\left( \mathrm{M} \right)
	+
	\sum\limits_{ \left( x  , y  \right) \in X \times Y \subseteq \mathbb{R}^2 } \left( \mathbf{K}\left( y \middle\vert \mathrm{M}\left( x \right)  \right) \right)^2
	\text{ ,}
\end{equation}
where 
$ X \times Y $ is given by the available dataset,
$ \lambda $ is the regularisation constant, and $ \mathbf{K}\left( \cdot \right) $ is the prefix algorithmic complexity.
By employing approximations to algorithmic complexity, the authors show that the property of differentiation is not required to achieve results similar to those obtained using differentiable programming such as deep learning.  

The chief advantage of such an approach is that these methods within the framework of AIT do not depend on mathematical constraints or conditions (such as differentiability), and are better equipped to deal with causal chains and compared to statistical approaches to machine learning.  
This is because: 
(1) it avoids overfitting on the available dataset (from which both the training set and the test set are selected) by design due to an inherent bias toward low-algorithmic-complexity models; 
and (2) can unveil deeper algorithmic structures not related to obvious redundancies, such as simple regularities that other entropic-like measures would miss and on which most loss functions are based.  
As shown in~\cite{Zenil2017a}, entropy is a weak measure of randomness that lacks the robustness of an invariance result as in AIT ~\cite{Calude2002,Chaitin2004,Downey2010,Li1997}.

Thus, it follows from these results the question whether or not assuming an underlying universal distribution as in the algorithmic coding theorem is sufficient for guaranteeing that generalisation or prediction would continue to be non over-fitted not only for the available data, but also for fresh data \cite{Witten2017} to which the learning process has no access prior to the estimation of the optimal model.
Unless one assumes other ad-hoc conditions (see for example Section~\ref{sectionComplexitycaging}), we shall see that the answer to this question is in general negative.

%\section{Learning algorithms, computability theory, and algorithmic information theory}\label{sectionAITandmachinelearning}
%\todo{resplit the into more subsections}

\section{Computability-theoretic machine learning}\label{sectionComputabilitytheoryandmachinelearning}

In this section, we focus on definitions of the concepts and mathematical tools used in machine learning so that these become based on computability-theoretic formalisation, and thus can be readily applicable to developing proofs using algorithmic information theory.
This is also important for highlighting the necessary and/or sufficient conditions in further on discussions about cases in which our results may not hold, thereby demanding future research. 
The objective of this article is to find the most general and abstract conditions for which our theorems hold, while machine learning methods and problems usually found in the literature remain as particular cases. 
This Section~\ref{sectionComputabilitytheoryandmachinelearning} has the purpose setting the ground for new lines of research concerning abstract theoretical investigations of machine learning problems that employs algorithmic information theory and computability theory.

\subsection{Data generating processes}\label{sectionDatageneratingprocess}

Let $ \mathrm{D} $ denote an arbitrary dataset.
In case of regression tasks, one has that $ \mathrm{D} = X_{ \mathrm{D} } \times Y $.
Our results are agnostic with respect to the data structure of the input dataset $  X_{ \mathrm{D} } $ or the target dataset $ Y $.
Learning algorithms, by its nature, as any other computational implementation that can receive information from external sources, can only take inputs from what can be encoded.
Thus, any scalar in any data point (which may be any sort of encodable multidimensional structure, such as a vector, a matrix or a network) of  $ \mathrm{D} $ may be assumed to be an encodable number that belongs to $ \mathbb{R} $ or $ \mathbb{C} $ without loss of generality.
%We only require both the sets $ \left\{ X_{ \mathrm{D} } \right\} $ and $ \left\{ Y \right\} $ of all possible input and target datasets, respectively, to be infinite computably enumerable; 
%%(and, as a consequence, that $ \left\{ \mathrm{D} \right\} $ is infinite computably enumerable) 
%and that there is an infinite computably enumerable set $ \mathbf{M} $ of models $ \mathrm{M} $ that satisfy\footnote{ Note that satisfying Definition~\ref{defModel} of a model does not automatically imply that this model needs to be optimal. A function $ \mathrm{M} $ can be a model of a regression task while performing poorly at approximating the actual function $ f $. See Definitions~\ref{defOptimality}, \ref{defOptimalmodelfromavailabledata}, and \ref{defGlobaloptima}.} Definition~\ref{defModel} with those $  X_{ \mathrm{D} } \in \left\{ X_{ \mathrm{D} } \right\} $ and $ Y \in \left\{ Y \right\} $.\todo{Move this part to the section about models.}
Note that the models in the machine learning literature that were mentioned in Section~\ref{sectionMachinelearning} (i.e., linear, logistic, and rectify) are examples of models (with their respective datasets ranging in $ \mathbb{R}^{ < \infty } $) that satisfy these conditions.

Any dataset $ \mathrm{D} $ that is somehow made available (e.g., by an external source as in Definitions~\ref{defDatageneratingsource} and~\ref{defUniversaldatageneratingprocess}) to the application of a chosen machine learning method is denoted by $ \mathrm{D}_a $.
The set of any \emph{fresh data} to which one has no access before the whole learning process ends is denoted by $ \mathrm{D}_{new} $.
Let $ \mathrm{D}_{ total } = \mathrm{D}_a \cup \mathrm{D}_{new} $ denote a dataset composed of the available dataset and the unavailable dataset.
For practical purposes in which the dataset $ \mathrm{D}_{ total } $ is encoded, one can define $ \mathrm{D}_{ total } $ as a composed sequence $ \mathrm{D}_{ total } = \left( \mathrm{D}_a , \mathrm{D}_{new} \right) $ without loss of generality.

We assume the condition that the set $ \left\{ \mathrm{D}_a \right\} $ of all possible available datasets must be infinite computably enumerable;
and from any particular $ \mathrm{D}_a $, the set $ \left\{ \mathrm{D}_{new} \right\} $ of all possible fresh data must be infinite computably enumerable.
This condition is for example satisfied for every infinite set $ \left\{ \mathrm{D}_{total} \right\} $ of encodable sets of real-valued vectors or matrices, as usually seen in machine learning problems.

Let $ t_{max} - t_{min} $ be the total computation running time of the implementation of the arbitrarily chosen machine learning method, which started at arbitrary time instant $ t_{min} $.
Let $ \mathcal{X}_{ \mathrm{D} } $ be an \emph{external source} that generates the datasets $ \mathrm{D}_a $ and $ \mathrm{D}_{new} $ such that it generates $ \mathrm{D}_a $ before $ t_{min} $ and generates $ \mathrm{D}_{new} $ after $ t_{max} $.
Formally:

\begin{definition}[Data generating source]\label{defDatageneratingsource}
	One says $ \mathcal{X}_{ \mathrm{D} } $ is a \emph{data generating source} to a computational implementation of a machine learning method starting at time instant $ t_{min} $ and finishing at time instant $ t_{max} $ 
	iff 
	$ \mathcal{X}_{ \mathrm{D} } \colon  \mathcal{T} \to  \mathcal{S} $ is a total function and the set $ \mathcal{S} $ is encodable by some program $ p $ such that
	\begin{equation}
		\mathrm{D}_a 
		=
		\left\{ \mathbf{U}\left( \left< \mathcal{X}_{ \mathrm{D} }\left( t \right) , p  \right> \right) \,\middle\vert\, t_0 \leq t < t_{min} \right\}
	\end{equation}
	and 
	\begin{equation}
		\mathrm{D}_{new} 
		=  
		\left\{ \mathbf{U}\left( \left< \mathcal{X}_{ \mathrm{D} }\left( t \right) , p  \right> \right) \,\middle\vert\, t_{max} < t < \infty  \right\}
		\text{ ,}
	\end{equation}
	where
	$ \mathcal{T} $ is the set of time instants (or indexes),
	$ \mathcal{S} $ is the set of states,
	$ \left\{ \mathbf{U}\left( \left< \mathcal{X}_{ \mathrm{D} }\left( t \right) , p  \right> \right) \,\middle\vert\, t_0 \leq t \leq \infty \right\} \subseteq \left\{ \mathrm{D} \right\} $,
	and $ \left< \cdot , \cdot \right> $ denotes the arbitrarily chosen recursive bijective pairing function \cite{Li1997,Downey2010}.
\end{definition}

Note that there is no assumption on the mechanisms or the nature (whether it is deterministic, stochastic, or mixed) of the underlying process that plays the role of the data generating source.
Thus, as usual, a \emph{data generating process} \cite{Goodfellow2016} of $ \mathrm{D}_{total} $ is defined by the sequence built from the indexing of the set 
$ \left\{ \mathbf{U}\left( \left< \mathcal{X}_{ \mathrm{D} }\left( t \right) , p  \right> \right) \,\middle\vert\, t_0 \leq t \leq t' < \infty \land t' \geq t_{max}  \right\} $.
In addition,
consistent with the usual definition of stochastic processes \cite{Cover2005},
if the data generating process is stochastic, then there is a probability space defined upon $ \mathcal{S} $ and the data generating process becomes a sequence of (stochastically) random variables whose respective outcomes are encoded.
A classical example is when the sequence 
$ \big( \mathbf{U}\left( \left< \mathcal{X}_{ \mathrm{D} }\left( t_0 \right) , p  \right> \right) , \mathbf{U}\left( \left< \mathcal{X}_{ \mathrm{D} }\left( t_0 + 1 \right) , p  \right> \right) , \dots , \mathbf{U}\left( \left< \mathcal{X}_{ \mathrm{D} }\left( t' \right) , p  \right> \right) \big) $ 
is in fact a sequence of encoded i.i.d. random variables $ \mathcal{X}_{ \mathrm{D} }\left( t_0 \right) $, 
$ \mathcal{X}_{ \mathrm{D} }\left( t_0 + 1 \right) $, $ \dots $, and $ \mathcal{X}_{ \mathrm{D} }\left( t' \right) $.
In this case, if the i.i.d. stochastic data generating process follows a uniform probability distribution, one already knows that the no free lunch theorem applies (see also Section~\ref{sectionMachinelearninginlargedatasets}). 
However, for the present purposes, we want to investigate the case in which there is an \emph{algorithmic-informational bias toward simplicity} in the data generating process.
Therefore,
one needs a particular type of data generating source (here denoted by $ \mathcal{X}_{ \mathbf{U} } $) whose associated data generating process follows the \emph{universal distribution} \cite{Calude2002} as in Equation~\ref{equationACT}.
Formally:

\begin{definition}[Universally distributed data generating process]\label{defUniversaldatageneratingprocess}
	We say a \emph{data generating process} 
	\begin{equation}\label{equationDefinitionUniversaldatageneratingprocess}
		\left\{ \mathbf{U}\left( \left< \mathcal{X}_{ \mathbf{U} }\left( t \right) , p  \right> \right) \,\middle\vert\, t_0 \leq t \leq t' < \infty \land t' \geq t_{max}  \right\}
	\end{equation}
	of the dataset $ \mathrm{D}_{total} $ is \emph{universally distributed}
	iff $ \mathcal{X}_{ \mathbf{U} } $ satisfies Definition~\ref{defDatageneratingsource}
	and
	there is a fixed constant $ C \in \mathbb{R}$ and a probability measure 
	$ \mathbf{P} \left[ \cdot \right] $ such that, for every $ \mathrm{D}_{total} $, 
	\begin{equation}\label{equationDefinitionUniversalprobability}
		\mathbf{P} \left[ \text{ ``dataset } \mathrm{D}_{total} \text{ occur''} \right] = 
		C \sum\limits_{ \mathbf{U}\left( p \right) 
			= \mathrm{D}_{total} } \frac{ 1 }{ 2^{ \left| p \right| } }
		\text{ .}
	\end{equation}
\end{definition}

Note that the constant $ C $ is only present in Equation~\ref{equationDefinitionUniversalprobability} to ensure that it is a probability measure and not a probability semi-measure, should not the prefix-free language $ \mathbf{L} $ of $ \mathbf{U} $ be complete (i.e., $ \sum\limits_{ p \in \mathbf{L} } \frac{ 1 }{ 2^{ \left| p \right| } } < 1 $) \cite{Abrahao2017publishednat}.
Otherwise, if the language is a complete code (i.e., $ \sum\limits_{ p \in \mathbf{L} } \frac{ 1 }{ 2^{ \left| p \right| } } = 1 $), then $ C = 1 $.\footnote{ Note that $ \sum\limits_{ \mathbf{U}\left( p \right) 
= \mathrm{D}_{total} } \frac{ 1 }{ 2^{ \left| p \right| } } \leq 1 $ always hold because $ \mathbf{U} $ is a prefix universal Turing machine.}

\subsection{The training, validation, and test sets}\label{sectionDatasets}

A learning algorithm (see Definition~\ref{defLearningalgorithm}) may employ computable procedures and calls to random number generators to manipulate, re-sample, and/or split this \emph{available dataset} $ \mathrm{D}_a $ in order to build the training, test, and validation sets (see also Section~\ref{sectionMachinelearning}).
In order to define these procedures, let $ \mathcal{X}^m $ be an arbitrary external source of events  (e.g., stochastic randomly generated numbers or pseudo-randomly generated numbers), events which are encoded in the form of a $m$-dimensional array (or vector) $ \mathbf{r} $.
One should not confuse this external source $ \mathcal{X}^m $ with the data generating source $ \mathcal{X}_{ \mathrm{D} } $ from Definitions~\ref{defDatageneratingsource} or~\ref{defUniversaldatageneratingprocess}, which is the one that generate the datasets $ \mathrm{D}_{total} $.
The source $ \mathcal{X}^m $ may be either a different source from, or the same as, $ \mathcal{X}_{ \mathrm{D} } $, but one to which the learning algorithm has also access during the learning process (i.e., between time instants $ t_{min} $ and $ t_{max} $).

In order to define the stage of building the training, test, and validation sets during a learning process, as usually done in machine learning methods, let 
\[
\begin{array}{lccc}
	f_{tr} \colon & \left\{ \mathrm{D}_a \right\} \times \mathcal{X}^m & \to & \left\{ \mathrm{D} \right\} \\
	& \left( \mathrm{D}_a , \mathbf{r} \right) & \mapsto & f_{tr}\left( \mathrm{D}_a , \mathbf{r} \right) = \mathrm{D}_{train}
\end{array} 
\]
be a total computable function that receives an available dataset along with externally generated events (encoded in $ \mathbf{r} $) as inputs; and returns a \emph{training set} $ \mathrm{D}_{train} $ as output.
Let 
\[
\begin{array}{lccc}
	f_{te} \colon & \left\{ \mathrm{D}_a \right\} \times \mathcal{X}^m & \to & \left\{ \mathrm{D} \right\} \\
	& \left( \mathrm{D}_a , \mathbf{r} \right) & \mapsto & f_{te}\left( \mathrm{D}_a , \mathbf{r} \right) = \mathrm{D}_{test}
\end{array} 
\]
be a total computable function that receives an available dataset along with externally generated events (encoded in $ \mathbf{r} $) as inputs; and returns a \emph{test set} $ \mathrm{D}_{test} $ as output.
Let 
\[
\begin{array}{lccc}
	f_{va} \colon & \left\{ \mathrm{D}_a \right\} \times \mathcal{X}^m & \to & \left\{ \mathrm{D} \right\} \\
	& \left( \mathrm{D}_a , \mathbf{r} \right) & \mapsto & f_{va}\left( \mathrm{D}_a , \mathbf{r} \right) = \mathrm{D}_{vali}
\end{array} 
\]
be a total computable function that receives an available dataset along with externally generated events (encoded in $ \mathbf{r} $) as inputs; and returns a \emph{validation set} $ \mathrm{D}_{vali} $ as output.
As we will see in Definition~\ref{defComputablelearningprocess}, our present scope is to study learning processes whose external random source $ \mathcal{X}^m $ to which they may call are in fact pseudo-random, and therefore the underlying generating process that defines the source $ \mathcal{X}^m $ is in fact computable.
Thus, for the present purposes, one can omit the presence of $ \mathcal{X}^m $ in the definitions of $ f_{tr} $, $ f_{te} $, and $ f_{va} $ without loss of generality.

Also note that in case the chosen machine learning method does only optimise the parameters from the raw input dataset $ \mathrm{D}_a $, if it only builds the training set and the test set but not the validation set, or if it only builds the training set but not the test set and the validation set, then Definitions~\ref{defLearningalgorithm} and~\ref{defLearningprocess} will still remain sound; and our results hold without loss of any generality.

\subsection{Performance measures and error functions}\label{sectionPerformancemeasure}

Since we are dealing with regression problems, then $ \mathrm{D} = X_{ \mathrm{D} } \times Y $. 
Therefore, the \emph{training error} is calculated by a \emph{training error function}
\begin{equation}\label{equationTrainingerrorfunction}
	\begin{array}{lccc}
		f_{\epsilon,tr} \colon & \mathbf{M} \times X_{ \mathrm{D}_{train} } \times Y_{train} & \to & Z_{train} \\
		& \left( \mathrm{M} , \mathbf{x} , \mathbf{y} \right) & \mapsto & f_{\epsilon,tr}\left( \mathrm{M} , \mathbf{x} , \mathbf{y} \right) = z
%		\text{ ,}
	\end{array}
\end{equation}
that is total and computable, where $ Z_{train} $ is a totally ordered set.
Analogously, we have the \emph{generalisation error} calculated by the \emph{test error function}
\begin{equation}\label{equationTesterrorfunction}
	\begin{array}{lccc}
		f_{\epsilon,te} \colon & \mathbf{M} \times X_{ \mathrm{D}_{test} } \times Y_{test} & \to & Z_{test} \\
		& \left( \mathrm{M} , \mathbf{x} , \mathbf{y} \right) & \mapsto & f_{\epsilon,te}\left( \mathrm{M} , \mathbf{x} , \mathbf{y} \right) = z
%		\text{ ,}
	\end{array}
\end{equation}
and the \emph{validation error} calculated by the \emph{validation error function}
\begin{equation}\label{equationValidationerrorfunction}
	\begin{array}{lccc}
		f_{\epsilon,va} \colon & \mathbf{M} \times X_{ \mathrm{D}_{vali} } \times Y_{vali} & \to & Z_{vali} \\
		& \left( \mathrm{M} , \mathbf{x} , \mathbf{y} \right) & \mapsto & f_{\epsilon,va}\left( \mathrm{M} , \mathbf{x} , \mathbf{y} \right) = z
	\end{array}
	\text{ .}
\end{equation}

Note that unlike function $ f_\epsilon $, here we also require these error functions to be computable. 
This is because, $ \mathrm{M} $ only needs the existence of an error function, which may be uncomputable, in order to the function $ \mathrm{M} $ be understood as a model.
However, learning algorithms (see Definition~\ref{defLearningalgorithm}) can only apply computable versions of (or computable approximations to) error functions.

For any arbitrarily chosen machine learning method, there is a criterion that some computable procedure tests with the purpose of deciding whether or not a model is optimal.
Thus, the functions, equations, inequalities, constants, etc that define this criterion must be embedded into the learning algorithm.
In the general case, there must be a \emph{performance measure} that evaluates a particular model $ \mathrm{M} $ with respect to the training, validation, and test errors from the respective training, valavidation, and test sets (or with respect to the error from the available dataset directly).
A simple example of optimality criterion is to check whether or not the 
MSEs calculated on the training set and the test set do not surpass certain thresholds.
Formally, there would be $ \epsilon_{tr} \in Z_{train}  \subseteq \mathbb{R} $  such that the training error in the following equation
\begin{equation}
	\text{MSE}_{tr}  
	= 
	\underset{ \scriptsize
	\begin{array}{c}
		\hat{ \mathbf{ y } } = \mathrm{M}\left( \mathbf{x} \right) \\
		\mathbf{x} \in X_{ \mathrm{D}_{train} } \\
		\mathbf{y} \in Y_{train}
	\end{array}
	}{ \mathbf{ E } }  \left[ \mathnormal{ \left\lVert \hat{ \mathbf{ y } } - \mathbf{ y } \right\rVert_{ l^2 }^2 } \right] 
	\leq 
	\epsilon_{tr}
	\text{ .}
\end{equation}
is expected to hold; 
and there would be $ \epsilon_{te} \in Z_{test} \subseteq \mathbb{R} $  such that the generalisation error in the following equation
\begin{equation}\label{equationMSEtesterror}
	\text{MSE}_{te}  
	= 
	\underset{ \scriptsize
	\begin{array}{c}
		\hat{ \mathbf{ y } } = \mathrm{M}\left( \mathbf{x} \right) \\
		\mathbf{x} \in X_{ \mathrm{D}_{test} } \\
		\mathbf{y} \in Y_{test}
	\end{array}
	}{ \mathbf{ E } }  \left[ \mathnormal{ \left\lVert \hat{ \mathbf{ y } } - \mathbf{ y } \right\rVert_{ l^2 }^2 } \right] 
	\leq
	\epsilon_{te}
%	\text{ .}
\end{equation}
is expected to hold.
However, in most practical cases, along with assuring the minimisation of the generalisation error below a certain level, more sophisticated trade-offs are employed, like setting a fine balance between bias in the test set and variance in the training set \cite{Goodfellow2016}.
Within the context of formal theories, computable functions, and decidability, this notion is formalised by:

\begin{definition}[Computability-theoretic performance measure]\label{defPerformancemeasure}
	Let $ \mathbf{F} $ be an arbitrary formal theory.
	Let 
	\begin{equation}\label{equationPerformancemeasure}
		\begin{array}{lccc}
			f_{per} \colon & \mathbf{M} \times \left\{ \mathrm{D}_{train} \right\} \times \left\{ \mathrm{D}_{validation} \right\} \times \left\{ \mathrm{D}_{test} \right\}  & \to & Z_{per} \\
			& \left( \mathrm{M} , \mathrm{D}_{train} , \mathrm{D}_{validation} , \mathrm{D}_{test} \right) & \mapsto &  z
		\end{array}
	\end{equation}
	be a function with $ \left| Z_{per} \right| \geq 2 $ such that $ \succeq $ defines a total order of $ Z_{per} $.
	Let $ Z_{ \neg opt } \subsetneq Z_{per} $ be a non-empty proper subset of $ Z_{per} $ that contains the non-optimal performance values, where the choice of $ Z_{ \neg opt } $ only depends on the formal theory $ \mathbf{F} $.
	We say $ f_{per} $  is a \emph{performance measure} according the formal theory $ \mathbf{F} $ iff: 
	\begin{itemize}
		\item for every $ z \in Z_{per} $ and $ z' \in Z_{ \neg opt } $ with $ z \succeq z' $, one has it that $ z \in Z_{ \neg opt } $;  
		
		\item and $ f_{per} $  is a total function that can be computed by some Turing machine when $ \mathbf{F} $ is given as its input.
	\end{itemize}
\end{definition}

Note that any performance measure $ f_{per} $ satisfying Definition~\ref{defPerformancemeasure} is a particular type of the error function $ f_{ \epsilon } $ in Definition~\ref{defModel}.

For practical purposes in machine learning problems, one should assume a type of ``ergodicity'' condition for the performance measure $ f_{per} $ so as to avoid a trivialisation of the problem of learning from the experience.
In other words, from any collection of data, there must be at least one larger collection of data, which contains the former, so that the performance measure on this larger collection has an arbitrary value $ z \in Z_{ \neg opt } \subsetneq Z_{per} $.
Otherwise, there might be some triple $ \left( \mathrm{D}_{train} , \mathrm{D}_{validation} , \mathrm{D}_{test} \right) $ from which any collection of fresh data, no matter how divergent from the former triple this new dataset is, will never be able to non-optimally perform.
In this case, it trivialises the learning problem because, even if the number of possible datasets is infinite and diverse, there might be a model and a special available dataset to which one can add an infinite and arbitrarily diverse amount of new data (possibly governed by a totally different model from the first one) without making the model become non optimal, or perform poorly.
Therefore, in order to avoid this trivialization in machine learning problems, we assume the condition that the performance measure $ f_{per} $ must measure a \emph{non-trivial} learning performance, i.e., for every $ \mathrm{D}_{train} , \mathrm{D}_{validation} , \mathrm{D}_{test} \subseteq \mathrm{D}_a $ and for every $ z \in Z_{ \neg opt } $, there are an infinite computably enumerable set $ \left\{ \mathrm{D}'_{test} \right\} $ of distinct datasets $ \mathrm{D}'_{test} \subseteq \mathrm{D}_{total} $ with $ f_{per}\left( \mathrm{M} , \mathrm{D}_{train} , \mathrm{D}_{validation} , \mathrm{D}'_{test} \right) \succeq z $ and $ \mathrm{D}_{test} \subsetneq \mathrm{D}'_{test} $.

Note that such a non-triviality condition for performance measures is satisfied by any machine learning method that, for example, employs a performance measure based on averaging errors (or distances) over the data points, e.g., the $ \text{MSE}_{te} $ in Equation~\ref{equationMSEtesterror}.
This is because no matter how low the value of $ \text{MSE}_{te} $ is when calculated on the available dataset $ \mathrm{D}_a $, there always is a sufficiently large number of distinct data points (in $ \mathrm{D}'_{test} $) that can be added in order to increase the value of 
\[
\underset{ \scriptsize
\begin{array}{c}
	\hat{ \mathbf{ y } } = \mathrm{M}\left( \mathbf{x} \right) \\
	\mathbf{x} \in X_{ \mathrm{D}'_{test} } \\
	\mathbf{y} \in Y'_{test}
\end{array}
}{ \mathbf{ E } }  \left[ \mathnormal{ \left\lVert \hat{ \mathbf{ y } } - \mathbf{ y } \right\rVert_{ l^2 }^2 } \right]
%\text{ .}  
\]
with respect to the former value of $ \text{MSE}_{te} $.

Also note that the conditions $ \left| Z_{per} \right| \geq 2 $ along with the non-empty $ Z_{ \neg opt } \subsetneq Z_{per} $ also avoid other trivialisations of the problem in cases which every model would be considered to be optimal, or alternatively every model would be considered to be non optimal.

Thus, these assumptions guarantee that there is always at least one $ \delta \in Z_{per} $ such that, for every $ x \succeq \delta $ with $ x \in Z_{per} $, one has it that: 
$ x \in Z_{ \neg opt } $; 
and there are an infinitely many $ \mathrm{D}'_{test} \subseteq \mathrm{D}_{total} $ with $ f_{per}\left( \mathrm{M} , \mathrm{D}_{train} , \mathrm{D}_{validation} , \mathrm{D}'_{test} \right) \succeq x $, even in the case $ f_{per}\left( \mathrm{M} , \mathrm{D}_{train} , \mathrm{D}_{validation} , \mathrm{D}_{test} \right) \preceq \delta $ holds.

Another assumption that one should make for most practical purposes in machine learning problems is for some type of ``stationarity'' condition for the performance measure.
This is necessary in order to guarantee that the optimal models are not constraining the size of the set of possible datasets for which these respective models are optimal to be finite, instead of infinite.
Otherwise, for a certain optimal model, one may have only a finite number of distinct available datasets for which this model is considered to be optimal by the performance measure.
Thus, one needs to assume that for every pair of model and available dataset for which the performance measure consider such a model to be optimal, there are infinitely many distinct available datasets for which this model continues to be deemed optimal by the performance measure.
Formally, we assume the condition that the performance measure $ f_{per} $ must also be \emph{extensible}, i.e., for every $ \mathrm{D}_a $ and for every $ \mathrm{M} $ with $ f_{per}\left( \mathrm{M} , \mathrm{D}_{train} , \mathrm{D}_{validation} , \mathrm{D}_{test} \right) \notin Z_{ \neg opt }  $, where $ \mathrm{D}_{train} , \mathrm{D}_{validation} , \mathrm{D}_{test} \subseteq \mathrm{D}_a $, 
there is an infinite computably enumerable set $ \left\{ \mathrm{D}_a \right\} $ of distinct available datasets $ \mathrm{D}'_a \in \left\{ \mathrm{D}_a \right\} $ such that $ f_{per}\left( \mathrm{M} , \mathrm{D}'_{train} , \mathrm{D}'_{validation} , \mathrm{D}'_{test} \right) \notin Z_{ \neg opt } $ holds, where $ \mathrm{D}'_{train} , \mathrm{D}'_{validation} , \mathrm{D}'_{test} \subseteq \mathrm{D}'_a $.

Like it was the case for the non-triviality condition, note that this extensibility condition for performance measures is satisfied by any machine learning method that, for example, employs a performance measure based on averaging errors (or distances) over the data points, e.g., the $ \text{MSE}_{te} $ in Equation~\ref{equationMSEtesterror}.
This occurs because one can always add distinct data points to construct a new larger dataset $ \mathrm{D}'_a $ (containing a new test set $ \mathrm{D}'_{test} $) so that 
\[
\underset{ \scriptsize
	\begin{array}{c}
	\hat{ \mathbf{ y } } = \mathrm{M}\left( \mathbf{x} \right) \\
	\mathbf{x} \in X_{ \mathrm{D}'_{test} } \\
	\mathbf{y} \in Y'_{test}
	\end{array}
}{ \mathbf{ E } }  \left[ \mathnormal{ \left\lVert \hat{ \mathbf{ y } } - \mathbf{ y } \right\rVert_{ l^2 }^2 } \right]
%\text{ .}  
\approx
\underset{ \scriptsize
	\begin{array}{c}
	\hat{ \mathbf{ y } } = \mathrm{M}\left( \mathbf{x} \right) \\
	\mathbf{x} \in X_{ \mathrm{D}_{test} } \\
	\mathbf{y} \in Y_{test}
	\end{array}
}{ \mathbf{ E } }  \left[ \mathnormal{ \left\lVert \hat{ \mathbf{ y } } - \mathbf{ y } \right\rVert_{ l^2 }^2 } \right]
\text{ .}
\]
For example, take the simplest case of a linear regression where $ \mathrm{M}\left( x \right) = a \, x + b $ is an optimal model for an available dataset $ \mathrm{D}_a $. Then, one can add an infinite number of distinct points in $ \mathbb{R}^2 $ so that the $ \text{MSE} $ with respect to the line given by equation $ \hat{ y } = a \, x + b $ is preserved.
The same also holds for the cross-entropy as in Equation~\ref{equationCrossentropy}: one can always add distinct data points so that the cross-entropy of KL divergence from the optimal probability distribution $ \hat{ y } $ is preserved.

\subsection{Models}\label{sectionModel}

First, we define a model (in particular for regression or prediction tasks) in machine learning.
The main idea formalised in Definiton~\ref{defModel} is that a model $ \mathrm{M} $ is an abstract form of the computably effective procedure that performs the task that the learning algorithm is trying to learn, including not only the parameters to be optimised, but also the hyperparameters (and, possibly any other procedure, data structure, formal theory, etc, that is embedded into the model).

We only require both the sets $ \left\{ X_{ \mathrm{D} } \right\} $ and $ \left\{ Y \right\} $ of all possible input and target datasets, respectively, to be infinite computably enumerable; 
%(and, as a consequence, that $ \left\{ \mathrm{D} \right\} $ is infinite computably enumerable) 
and that there is an infinite computably enumerable set $ \mathbf{M} $ of models $ \mathrm{M} $ that satisfy\footnote{ Note that satisfying Definition~\ref{defModel} of a model does not automatically imply that this model needs to be optimal. A function $ \mathrm{M} $ can be a model of a regression task while performing poorly at approximating the actual function $ f $. See Definitions~\ref{defOptimality}, \ref{defOptimalmodelfromavailabledata}, and \ref{defGlobaloptima}.} Definition~\ref{defModel} with those $  X_{ \mathrm{D} } \in \left\{ X_{ \mathrm{D} } \right\} $ and $ Y \in \left\{ Y \right\} $.

\begin{definition}[Model for regression tasks]\label{defModel}
	We say 
	\begin{equation}\label{equationDefinitionModel}
	\begin{array}{lccc}
	\mathrm{M} \colon & X_{ \mathrm{D} } \subseteq X & \to & Y \\
	&  \mathbf{x} & \mapsto & \mathrm{M}\left( \mathbf{x} \right) = \hat{ \mathbf{y} }
	\end{array}
	\end{equation} 
	is a \emph{model} of a function $ f $
	iff: 
	\begin{itemize}
		\item $ \mathrm{M} $ is a partial computable function;
		
		\item $ X_{ \mathrm{D} } $ is an infinite computably enumerable set; 
		
		\item 
		\begin{equation}\label{equationDefinitionTruefunction}
		\begin{array}{lccc}
		f \colon & X & \to & Y \\
		&  \mathbf{x} & \mapsto & f\left( \mathbf{x} \right) = \mathbf{y}
		\end{array}
		\end{equation}
		is a total function when its domain is constrained to $ X_{ \mathrm{D} } \subseteq X $;
		
		\item there is a non-empty set $ \mathbf{M} $ and there is at least one total function (i.e., the \emph{error function}) 
		\begin{equation}\label{equationDefinitionErrorfunction}
		\begin{array}{lccc}
		f_{\epsilon} \colon & \mathbf{M} \times X_{ \mathrm{D} } \times Y & \to & Z \\
		& \left( \mathrm{M} , \mathbf{x} , f\left( \mathbf{x} \right) \right) & \mapsto & f_{\epsilon}\left( \mathrm{M} , \mathbf{x} , f\left( \mathbf{x} \right) \right) = z
		%		\text{ ,}
		\end{array}
		\end{equation}
		such that $ Z $ is a totally ordered set and $ M \in \mathbf{M} $.
	\end{itemize}

\end{definition}

%\color{blue}\todo{Update}
%For example, in the first Definition 1 there is a condition (line 318) that is always true if understood literally: there is always a non-empty set $\mathbf{M}$ and at least one total function $f_\epsilon$, just take $\mathbf{M}=\{\mathrm{M}\}$, let $Z$ be a singleton and let $f_\epsilon$ be the only function to $Z$. Probably the authors had something else in mind (e.g., that the notion of model they define include the family $\mathbf{M}$ and function $f_\epsilon$), but then no conditions on these objects are included in the definition. The formula (10) is also confusing: is $\mathrm{M}$
%the same function that is mentioned earlier, or it is just a free variable? What is meant by $f(\mathbf{x})$ in the argument of the function? 
%
%\color{green}
%
%Definition 3.1 talks about the function M defined on abstract sets X
%and Y, and has some computability properties. Without knowing the structure of these
%sets, this definition does not make any sense. The same issues appear in most definitions,
%so proving results is impossible at this stage.
%
%
%\color{black}

In case the model is said to have parameters and hyperparameters, then we have it that $ \mathrm{M}\left( \mathbf{ x } \right) = f_{ \mathrm{M} }\left( \boldsymbol{ \theta } , \pmb{ \gamma } , \mathbf{ x } \right) $, where: 
\begin{equation}\label{equationDefinitionFunctionfromparameters}
\begin{array}{lccc}
f_{ \mathrm{M} } \colon & X_{ \mathrm{D} } \times \Theta \times \Gamma & \to & Y \\
&  \left( \mathbf{x} , \boldsymbol{ \theta } , \pmb{ \gamma } \right) & \mapsto & f_{ \mathrm{M} }\left( \mathbf{x} , \boldsymbol{ \theta } , \pmb{ \gamma } \right) = \hat{ \mathbf{y} }
\end{array}
\end{equation}
is a partial computable function;
$ \boldsymbol{ \theta } = \left( \theta_1, \theta_2 , \dots , \theta_n , \dots \right) \in \Theta  $ are the parameters of the model $ \mathrm{M} $; 
and $ \pmb{ \gamma } = \left( \gamma_1, \gamma_2 , \dots , \gamma_{n'} \right) \in \Gamma $ are the hyperparameters of the same model $ \mathrm{M} $.
%\footnote{ Note that $ \omega $ here denotes the first infinite ordinal number.}
%and both $ \Theta $ and $ \mathrm{X}_{ \pmb{ \gamma } }^{ < \omega } $ are infinite.
Otherwise, if the model only has parameters, then $ \mathrm{M}\left( \mathbf{ x } \right) = f_{ \mathrm{M} }\left( \boldsymbol{ \theta } , \mathbf{ x } \right) $.
For the sake of computability-theoretic purposes, we can conflate both parameters and hyperparameters into just one set and call it the set of parameters from now on without loss of generality.

When 
\[ \max\left\{ \left| \boldsymbol{ \theta } \right| \, \big\vert \,    \boldsymbol{ \theta } \in \Theta \right\} \in \mathbb{N} \text{ ,} \]
then the model $ \mathrm{M} $ 
is called \emph{parametric}.
Otherwise, when there may be an infinite number of parameters or when the number of parameters grows with the available dataset size, it is called \emph{non-parametric}.
%{\color{blue} We shall see in Lemma~\ref{label} that Turing machines are non-parametric models.}\todo{Check if this result is indeed necessary.}

Note that  $ f\left( \cdot \right) $ is the actual function the model $ \mathrm{M} $ is trying to approximate.
Also note that particularly in Definition~\ref{defModel}, there is no obligation:
for the function $ \mathrm{M} $ to be total on domain $ X_{ \mathrm{D} } $;
and for the function $ f_{\epsilon} $ to be computable.
%These will be only requirements in Definitions~\ref{defLearningalgorithm} and~\ref{defLearningprocess} of learning algorithms and learning processes, respectively.

It is easy to see that the models mentioned in Section~\ref{sectionMachinelearning} (e.g., the ones for linear regression, logistic regression, ReLU, and ANNs) satisfy Definition~\ref{defModel} with any of the error functions also mentioned in Section~\ref{sectionMachinelearning} (e.g., MSE or cross-entropy).
%In fact, since applications of machine learning methods are empirical implementations of effectively computational methods---regardless of other mathematical conditions such as differentiability, which only affect the fact whether or not the learning algorithm is able to find an optimal model---, 
%one can always find a function $ f_{\epsilon} $ for the computable procedure of calculating the error (or loss) that the chosen machine learning method is implementing.
%In addition, one can always find a function $ f $ for the actual computable procedure that performs the regression task that the respective machine learning method is implementing.
%Thus, for any effective and computationally feasible machine learning method, there is always a way to choose functions $ f $ and $ f_{\epsilon} $ so that the algorithm or procedure performing the task (i.e., the prediction) satisfies Definition~\ref{defModel}.
For example, even in the case of uncomputable error functions such as in Equation~\ref{equationRegularisedalgorithmiclossfunction}, where computable upper bounds for $ \mathbf{K}\left( \cdot \right)$ are employed to approximate the value of $ J_{ K } $, we have it that there is $ f_{\epsilon} $ satisfying Definition~\ref{defModel}, in particular when $ f_{\epsilon} = J_{ K } $.
Note that, in this example, $ f_{\epsilon} = J_{ K } $ is a semi-computable function.

\subsection{Optimality}\label{sectionOptimality}

One key idea in most machine learning methods in the literature is that a model is considered to be \emph{optimal} iff one decides that some arbitrarily chosen trade-off between test error and training error (and/or the bias-variance trade-off) and any other hyperparameter optimisation criteria (such as in regularisation) \cite{Goodfellow2016,Witten2017} are satisfied by a given model.
From a given performance measure, this is formalised by:

\begin{definition}[Computability-theoretic optimality criterion]\label{defOptimality}
	Let $ f_{per} $  be a \emph{performance measure} according the formal theory $ \mathbf{F} $ that satisfy Definition~\ref{defPerformancemeasure}
	Let $ p_{opt} $ be a program of a (total) Turing machine such that, for every $ \left( \mathrm{M} , \mathrm{D}_{train} , \mathrm{D}_{validation} , \mathrm{D}_{test} \right) $, the following holds
		\begin{equation}
			\begin{aligned}
				\mathbf{U}\left( \left< \mathbf{F} , \mathrm{M} , \mathrm{D}_{train} , \mathrm{D}_{validation} , \mathrm{D}_{test} , p_{opt} \right> \right) \\
				= 
				\begin{cases}
					1 
					& \textnormal{, if $ f_{per}\left( \mathrm{M} , \mathrm{D}_{train} , \mathrm{D}_{validation} , \mathrm{D}_{test} \right) \notin Z_{ \neg opt } $ } \\
					0
					& \textnormal{, if $ f_{per}\left( \mathrm{M} , \mathrm{D}_{train} , \mathrm{D}_{validation} , \mathrm{D}_{test} \right) \in Z_{ \neg opt } $ }
				\end{cases}
				\text{ .}
			\end{aligned}
		\end{equation}
	We say $ \mathrm{M} $ is \emph{optimal} according to the performance measure $ f_{per} $ and the formal theory $ \mathbf{F} $ iff
	$ \mathbf{U}\left( \left< \mathbf{F} , \mathrm{M} , \mathrm{D}_{train} , \mathrm{D}_{validation} , \mathrm{D}_{test} , p_{opt} \right> \right) 
	=
	1  $.
\end{definition}

As mentioned to the above, in cases the training, validation, and/or test sets are omitted or are not built, note that one can define the performance measure $ f_{per} $ and the optimality computable criterion $ p_{opt} $ with 
$ f_{per} \colon  \mathbf{M} \times \left\{ \mathrm{D}_a \right\}   \to  Z_{per} $ and $ \mathbf{U}\left( \left< \mathbf{F} , \mathrm{M} , \mathrm{D}_a , p_{opt} \right> \right) = 1  $ without any loss of generality in our forthcoming results.

\subsection{Learning processes}\label{sectionLearningprocesses}

The key idea of $ \mathrm{P} $ being a learning algorithm for regression problems is that: 
it is a total computable function from the external sources; 
and it includes the learner, the optimisation algorithm, and any other procedure that ends up returning the model, final parameters and/or hyperparameters from the available data.
Thus, the algorithms for sampling and retrieving the training set, test set and validation set, calculating the loss/cost/error functions, applying gradient descent, cross-validation, regularisation, dropouts, early stopping, bias-variance trade-off, hyperparameter tunning, etc are already embedded into $ \mathrm{P} $.
In summary, $ \mathrm{P} $ implements a fully automated learning process without external (human) intervention.
%That is, the whole learning process is computable.
Formally:

\begin{definition}[Learning algorithm]\label{defLearningalgorithm}
	Let $ \mathcal{X}_{ \mathrm{D} } $ be an arbitrary data generating source that satisfies Definition~\ref{defDatageneratingsource}.
	Let $ f_{per} $ be an arbitrary performance measure that satisfies Definition~\ref{defOptimality} with a formal theory $ \mathbf{F} $.
	Let $ \mathcal{X}^m $ be an external source of events.
	We say that
	\begin{equation}
		\begin{array}{lccc}
			\mathrm{P} \colon & \left\{ \mathrm{D}_a \right\} \times \mathcal{X}^{ m \times \left( t_{max} - t_{min} \right) } & \to & \mathbf{M} \times \left\{ 0 , 1 \right\} \times Z_{per} \\
			& \left( \mathrm{D}_a , \mathbf{r} \right) & \mapsto & \mathrm{P}\left( \mathrm{D}_a , \mathbf{r} \right) = \left( \mathrm{M} , \mathbf{U}\left( \left< \mathbf{F} , \mathrm{M} , \mathrm{D}_a , p_{opt} \right> \right) ,
			f_{per}\left( \mathrm{M} , \mathrm{D}_a \right)
			\right)
		\end{array} 
	\end{equation}
	is a \emph{learning algorithm} according to the performance measure $ f_{per} $ and formal theory $ \mathbf{F} $ iff
	it is a total computable function 
%	and
%	$ \mathbf{U}\left( \left< \mathbf{F} , \mathrm{M} , \mathrm{D}_a , p_{opt} \right> \right) = 1 $,
	where any $ \mathrm{D}_a $ is generated by $ \mathcal{X}_{ \mathrm{D} } $
	and any model $ \mathrm{M} \in \mathbf{M} $ satisfies Definition~\ref{defModel}.
\end{definition}

%Note that the error function $ f_{ \epsilon } $ in Definition~\ref{defModel} become just a particular type of a performance measure $ f_{per} $ in case $ f_{ \epsilon } $ is computable and

As usual, let $ \mathbf{U}_t\left(  \mathrm{D}_a , r_t , \mathrm{P}  \right) $ denote the configuration of the machine $ \mathbf{U} $ at time instant $ t $ when running on input $ \left<  \mathrm{D}_a , r_t , \mathrm{P}  \right> $, where $ r_t = \mathcal{X}^m\left( t \right) $ and $ \mathbf{r} = \left( r_{ t_{min} } , ... , r_{ t_{max} } \right) $.
Thus, note that $ \mathbf{U}_{ t_{max} }\left(  \mathrm{D}_a , r_{ t_{max} } , \mathrm{P} \right) $ is the last configuration that corresponds to the halting state of machine $ \mathbf{U} $ with input $ \left<  \mathrm{D}_a , r_t , \mathrm{P}  \right> $, returning $ \left( \mathrm{M} , \mathbf{U}\left( \left< \mathbf{F} , \mathrm{M} , \mathrm{D}_a , p_{opt} \right> \right) ,
f_{per}\left( \mathrm{M} , \mathrm{D}_a \right)
\right) $ as its output.
In this way, by combining the data generating process from Definition~\ref{defDatageneratingsource} and the computational implementation of the learning algorithm in Definition~\ref{defLearningalgorithm} we define a \emph{machine learning process} by:

\begin{definition}[Learning process]\label{defLearningprocess}
	Let $ \mathrm{P} $ be a learning algorithm satisfying Definition~\ref{defLearningalgorithm} with the data generating source $ \mathcal{X}_{ \mathrm{D} } $, external source $ \mathcal{X}^m $, performance measure $ f_{per} $, and formal theory $ \mathbf{F} $.
	Let $ \mathcal{L}\left( \mathcal{X}_{ \mathrm{D} } , \mathcal{X}^m , f_{per} , \mathbf{F} , \mathrm{P} \right) $ denote the sequence
	\begin{equation}
		\begin{aligned}
		\Big( 
		\mathbf{U}\left( \left< \mathcal{X}_{ \mathrm{D} }\left( t_0 \right) , p  \right> \right) , \dots , \mathbf{U}\left( \left< \mathcal{X}_{ \mathrm{D} }\left( t_{min} - 1  \right) , p  \right> \right) , \\ 
		\mathbf{U}_{ t_{min} }\left(  \mathrm{D}_a , r_{ t_{min} } , \mathrm{P}  \right)  , \dots ,
		\mathbf{U}_{ t_{max} }\left(  \mathrm{D}_a , r_{ t_{max} } , \mathrm{P} \right) , \\
		\mathbf{U}\left( \left< \mathcal{X}_{ \mathrm{D} }\left( t_{max} + 1 \right) , p  \right> \right) , \dots , \mathbf{U}\left( \left< \mathcal{X}_{ \mathrm{D} }\left( t'  \right) , p  \right> \right)
		\Big)
		\text{ .}
		\end{aligned}
	\end{equation}
	Then, we say $ \mathcal{L}\left( \mathcal{X}_{ \mathrm{D} } , \mathcal{X}^m , f_{per} , \mathbf{F} , \mathrm{P} \right) $
	is a \emph{learning process} of the learning algorithm $ \mathrm{P} $ from the data generating source  $ \mathcal{X}_{ \mathrm{D} } $.
\end{definition}

Note that the presence of the external source $ \mathcal{X}^m $ formalises the learning algorithm making use of re-sampling, randomised dropouts, noise introduction, or any other call to random trials in order to estimate the optimal model for a given available dataset.
%In this case, we say the learning process is \emph{mixed}, i.e., partially computable and partially stochastic.
%Otherwise, if there is no call to stochastic events in the learning algorithm, then we say the learning process is \emph{computable}.
If the random events generators called by the learning algorithm are actually pseudo-random instead of truly stochastically random, then the learning process is still computable:

\begin{definition}[Computable learning process]\label{defComputablelearningprocess}
	We say $ \mathcal{L}\left( \mathcal{X}_{ \mathrm{D} } , \mathcal{X}^m , f_{per} , \mathbf{F} , \mathrm{P} \right) $ is a \emph{computable learning process} iff
	$ \mathcal{L}\left( \mathcal{X}_{ \mathrm{D} } , \mathcal{X}^m , f_{per} , \mathbf{F} , \mathrm{P} \right) $ satisfies Definition~\ref{defLearningprocess} and
	the middle segment $ \big( 
	\mathbf{U}_{ t_{min} }\left(  \mathrm{D}_a , r_{ t_{min} } , \mathrm{P} \right)  , \dots , 
	\mathbf{U}_{ t_{max} }\left(  \mathrm{D}_a , r_{ t_{max} } , \mathrm{P} \right) %, \\
%	\mathbf{U}\left( \left< \mathcal{X}_{ \mathrm{D} }\left( t_{max} + 1 \right) , p  \right> \right) ,  \dots , \mathbf{U}\left( \left< \mathcal{X}_{ \mathrm{D} }\left( t'  \right) , p  \right> \right) 
	\big)  
	$
	is computable from any initial segment 
	$ \big( \mathbf{U}\left( \left< \mathcal{X}_{ \mathrm{D} }\left( t_0 \right) , p  \right> \right) , 
	\dots , \mathbf{U}\left( \left< \mathcal{X}_{ \mathrm{D} }\left( t_{min} - 1  \right) , p  \right> \right) \big) $.
\end{definition}

Note that a computable learning process does not imply that the data generating process of $ \mathcal{X}_{ \mathrm{D} } $ is computable.
It only implies that $ \mathcal{X}^m $ has an underlying process that is computable.
Thus, if the learning process is computable, one can already include within $ \mathrm{P} $ the underlying computable process that generates the pseudo-random events without loss of generality in our forthcoming results.
In this latter case, the algorithmic complexity of $ \mathrm{P} $ shall already encompass not only the algorithmic information content of the original learning algorithm, but also the algorithmic information content of the computable process that generates the pseudo-random events in $ \mathcal{X}^m $.
In this article, we only deal with learning processes that are computable with or without access to pseudo-random generators.
The investigation of the present results in the context of learning process in which there are calls to truly stochastically random events is an interesting open problem.
For example, this is tackled in \cite{Colbrook2022} for particular types of ANNs and available training data.
Thus, whether or not our results can be extended to the case in which $ \mathcal{X}^m $ is a stochastically random source is a necessary future research.

Any condition stating that the learning process is \emph{actually} acquiring better performance in finding an optimal model from the experienced data (which is generated by $ \mathcal{X}_{ \mathrm{D} } $) lies beyond the scope of the present article.
Since machine learning is an empirical problem (due to the presence of an external data generating source) in addition to a computability-theoretic problem of optimization, modeling, and prediction, the soundness or correctness of the performance measure and the computable optimality criterion $ p_{opt} $ with respect to the actual function $ f $ (see Definition~\ref{defModel}), which possibly models a real-world (empirical) unknown data generating source $ \mathcal{X}_{ \mathrm{D} } $, are in turn totally dependent on the chosen formal theory $ \mathbf{F} $.
For the present purposes of investigating the theoretical limits of machine learning, one can only study the mathematical properties of a learning algorithm instantiated as a computable process that receives inputs from $ \mathcal{X}_{ \mathrm{D} } $ (and $ \mathcal{X}^m $).
Whether or not the formal theory and the performance measure are sound and correct is an empirical problem (or, at least an inexorably ad-hoc condition) that reflects how well the chosen formal theory (along with the performance measure) is capable of formalising the correct intuition of how much ``$ \mathrm{M} $ is closer to (or approximates) $ f $'',
and thus it is a problem that lies beyond the scope of theoretical computer science.
For our pure theoretical purposes, one can assume that the chosen formal theory $ \mathbf{F} $ is sound and correct with respect to the experience without loss of generality.

\subsection{Learning optimal models from available data}\label{sectionOptimalmodels}

In this way, we now formalise the notion that “A computer program is said to learn from experience E with respect to some class of tasks T and performance measure P, if its performance at tasks in T, as measured by P, improves with experience E” \cite{Goodfellow2016,Mitchell1997}, according to the chosen performace measure and formal theory.
This formalisation of the learning algorithm being able to find a locally (or, possibly, globally) optimal model straightforwardly derives from Definitions~\ref{defOptimality} and~\ref{defLearningalgorithm} that are already embedded into Definition~\ref{defLearningprocess}:

\begin{definition}[Optimal model from available data]\label{defOptimalmodelfromavailabledata}
	Let $ \mathcal{L}\left( \mathcal{X}_{ \mathrm{D} } , \mathcal{X}^m , f_{per} , \mathbf{F} , \mathrm{P} \right) $ be a learning process that satisfies Definition~\ref{defLearningprocess}, where
	\begin{equation}
		\mathrm{D}_a 
		=
		\big( \mathbf{U}\left( \left< \mathcal{X}_{ \mathrm{D} }\left( t_0 \right) , p  \right> \right) , \\ 
			\dots , \mathbf{U}\left( \left< \mathcal{X}_{ \mathrm{D} }\left( t_{min} - 1  \right) , p  \right> \right) \big) 
	\end{equation}
	We say $ \mathrm{M} $ is an \emph{optimal model} from the available dataset $ \mathrm{D}_a $ according to learning algorithm $ \mathrm{P} $ iff
	$ \mathrm{P}\left( \mathrm{D}_a , \mathbf{r} \right) = \left( \mathrm{M} , 1 , z \right) $, where $ z \notin Z_{ \neg opt } $.
\end{definition}

\begin{notation}\label{defModelcalculatedbylearningalgorithm}
	Let $ \mathrm{M}_{ \left( \mathrm{P} , \mathrm{D}_a \right) } $ denote the \emph{optimal} model satisfying Definition~\ref{defOptimalmodelfromavailabledata} that was \emph{calculated} by an arbitrary learning algorithm $ \mathrm{P} $ from an arbitrary available dataset $ \mathrm{D}_a $.
\end{notation}

Note that for the sake of simplifying our proofs, Definition~\ref{defLearningalgorithm} of learning algorithms and Definition~\ref{defOptimalmodelfromavailabledata} optimal models were based on a single model that is the output of $ \mathrm{P} $.
Nevertheless, our forthcoming results hold without loss of generality if instead of a sole model, Notation~\ref{defModelcalculatedbylearningalgorithm} of $ \mathrm{M}_{ \left( \mathrm{P} , \mathrm{D}_a \right) } $ represents an encoded class 
$ \left( \mathrm{M}_1 , \dots , \mathrm{M}_n \right)  $
of optimal models such that the alternative form of Definition~\ref{defOptimalmodelfromavailabledata} becomes stated for $ \mathrm{P}\left( \mathrm{D}_a , \mathbf{r} \right) = \left( \mathrm{M}_{ \left( \mathrm{P} , \mathrm{D}_a \right) } , \mathbf{ 1 } , \mathbf{ z } \right) $, where $ n $ depends only on $ \mathrm{P} $ and for every $ \mathrm{M}_i \in \mathrm{M}_{ \left( \mathrm{P} , \mathrm{D}_a \right) } $, one has it that $ f_{per}\left( \mathrm{M}_i , \mathrm{D}_a \right) = z_i \notin Z_{ \neg opt } $ and
$ \mathbf{U}\left( \left< \mathbf{F} , \mathrm{M}_i , \mathrm{D}_a , p_{opt} \right> \right) = 1 $ hold.

%In this article, we only study the learning algorithms $ \mathrm{P} $ that can always return an optimal model from any given dataset generated from $ \mathcal{X}_{ \mathrm{D} } $.
%The investigation of learning algorithms that cannot find locally optimum models is beyond of the scope of the present article.
%Therefore, together with the non-trivialities conditions of the performance measure that ensure the existence of the lower bound $ \delta \in Z_{per} $ (see Section~\ref{sectionPerformancemeasure}), we guarantee that the number of distinct optimal models that a learning algorithm can return is infinite computably enumerable.

The formalisation of global optima straightforwardly follows from Definitions~\ref{defOptimality} and~\ref{defLearningprocess}:

\begin{definition}[Globally optimal model]\label{defGlobaloptima}
	Let $ \mathrm{M} $ be an \emph{optimal model} that satisfies Definition~\ref{defOptimalmodelfromavailabledata} with
	$ \mathcal{L}\left( \mathcal{X}_{ \mathrm{D} } , \mathcal{X}^m , f_{per} , \mathbf{F} , \mathrm{P} \right) $ and
	\begin{equation}
		\mathrm{D}_a 
		=
		\big( \mathbf{U}\left( \left< \mathcal{X}_{ \mathrm{D} }\left( t_0 \right) , p  \right> \right) , \\ 
		\dots , \mathbf{U}\left( \left< \mathcal{X}_{ \mathrm{D} }\left( t_{min} - 1  \right) , p  \right> \right) \big) 
		\text{ .}
	\end{equation}
	Let
	\begin{equation}
		\mathrm{D}_{new} 
		=
		\big( \mathbf{U}\left( \left< \mathcal{X}_{ \mathrm{D} }\left( t_{max} + 1 \right) , p  \right> \right) , \dots , \mathbf{U}\left( \left< \mathcal{X}_{ \mathrm{D} }\left( t'  \right) , p  \right> \right) \big) 
	\end{equation}
	and
	\begin{equation}
		\mathrm{D}_{total} 
		=
		\big( \mathrm{D}_a , \mathrm{D}_{new} \big) 
		\text{ .}
	\end{equation}
	We say $ \mathrm{M} $ is a \emph{globally optimal model} of the data generating source $ \mathcal{X}_{ \mathrm{D} } $ according to the performance measure $ f_{per} $ and formal theory $ \mathbf{F} $ iff
	$  \mathbf{U}\left( \left< \mathbf{F} , \mathrm{M} , \mathrm{D}_{total}  , p_{opt} \right> \right) = 1   $.
\end{definition}

%First, we prove in Lemma~\ref{label} that there always are data generating mechanisms that are complex enough to generate a dataset whose algorithmic complexity is as large as one wishes:
%
%\begin{lemma}
%	Let $ c \in \mathbb{N} $ be an arbitrary constant.
%	Let $ \mathrm{P} $ be an arbitrary learning algorithm.
%	Then, there is a dataset $ \mathrm{D} $ such that $ \mathbf{K}\left( \mathrm{P} \right) + c < \mathbf{K}\left( \mathrm{D} \right) $.
%\end{lemma}

\section{Deceivability of datasets and unpredictability of models from algorithmic randomness}\label{sectionRandomnessandmachinelearning}

%Sections~\ref{sectionModel} to~\ref{sectionOptimalmodels} give a computability-theoretic foundation for machine learning according to which was briefly presented in Section~\ref{sectionMachinelearning}. 
The present section is predicated upon the concepts formalised in the previous Section~\ref{sectionComputabilitytheoryandmachinelearning}, which give a computability-theoretic foundation for machine learning according to which was briefly presented in Section~\ref{sectionMachinelearning}.
Now, we introduce two new properties of learning processes that are closely related to algorithmic information theory: unpredictable models and unpredictably deceiving datasets.

We follow the usual notion of unpredictability from algorithmic randomness that demonstrates the equivalence between incompressibility, Martin-Löf tests, and martingales \cite{Downey2010,Calude2002,Chaitin2004} for infinite sequences.
Remember that a real number (or infinite binary sequence) $ w \in \left[ 0 , 1 \right] \subset \mathbb{R} $ is said to be \emph{algorithmically random} ($1$-random, K-random, or prefix algorithmically random) \cite{Chaitin2004,Calude2002,Downey2010} iff it satisfies
\[
\mathbf{K}\left( w \upharpoonright_n \right) \geq n - \mathbf{O}(1)
\text{ ,}
\]
\noindent where $ n \in \mathbb{N} $ is arbitrary and $ w \upharpoonright_n $ denotes the sequence of length $n$ corresponding to the first $ n $ fractional bits of the real number $ w $.

Within the context of finite objects,
a finite string $ y $ can be understood as \emph{unpredictable} from the finite string $ x $ by a machine $ \mathbf{U} $ and a formal theory $ \mathbf{F} $ iff there is a fixed constant $ C < \mathbf{K}\left( y \right) $ (which only depends on the chosen machine and formal theory, but not on the objects $ x $ or $ y $) such that 
$ \mathbf{K}\left( y \, \middle\vert \, \left< x, \mathbf{F} \right>  \right) 
\geq 
\mathbf{K}\left( y \right) - C $.
%where 
%$ xy $ denotes the concatenation of these two strings and 
%$ \left| \cdot \right| $ denotes the length of the string.
This definition roughly guarantees that any program (running on machine $ \mathbf{U} $ with formal theory $ \mathbf{F} $ with input) needs at least one extra bit of algorithmic information to retrieve the sequence $y$ from the sequence $ x $.
More formally, as $ \mathbf{K}\left( y \right) $ increases in comparison to $ \left| x \right| $ and $ \mathbf{K}\left( \mathbf{F} \right) $, any program will approximately need $ \mathbf{K}\left( y \right) $ bits of algorithmic information in order to compute $y$ from the sequence $ x $, i.e.,
\begin{equation*}
	\begin{aligned}
		\left| x \right| + \mathbf{K}\left( \mathbf{F} \right) = \mathbf{o}\left( \mathbf{K}\left( y \right) \right)
	\end{aligned}
\end{equation*} 
implies that 
\begin{equation*}
	\begin{aligned}
		\mathbf{K}\left( x \right) + \mathbf{K}\left( \mathbf{F} \right) = \mathbf{o}\left( \mathbf{K}\left( y \, \middle\vert \, \left< x, \mathbf{F} \right>  \right) \right)
%		\text{ ,}
	\end{aligned}
\end{equation*}
and
\begin{equation*}
	\begin{aligned}
		\mathbf{K}\left( y \, \middle\vert \, \left< x, \mathbf{F} \right>  \right)=
		\mathbf{ \Theta }\left( \mathbf{K}\left( y \right) \right)
		\text{ ,}
	\end{aligned}
\end{equation*}
%where $ f(x)=\mathbf{o}( g(x) ) $ denotes the usual strong asymptotic dominance (little-$ \mathbf{o} $ notation) when an arbitrary function $g$ dominates an arbitrary function $f$.
where: 
$ \mathbf{o}( \cdot ) $ denotes the little-$ \mathbf{o} $ notation for strong asymptotic dominance from above;
and $ \mathbf{ \Theta }( \cdot ) $ denotes the Big-$ \mathbf{ \Theta } $ notation for weak asymptotic dominance from above and below.

%This is in consonance with algorithmic randomness theory:
%remember that a real number (or infinite binary sequence) $ w \in \left[ 0 , 1 \right] \subset \mathbb{R} $ is said to be \emph{algorithmically random} ($1$-random, K-random, or prefix algorithmically random) \cite{Chaitin2004,Calude2002,Downey2010} iff it satisfies
%\[
%\mathbf{K}\left( w \upharpoonright_n \right) \geq n - \mathbf{O}(1)
%\text{ ,}
%\]
%\noindent where $ n \in \mathbb{N} $ is arbitrary and $ w \upharpoonright_n $ denotes the sequence of length $n$ corresponding to the first $ n $ fractional bits of the real number $ w $.
Therefore, note that for any particular machine $ \mathbf{U} $, any particular formal theory $ \mathbf{F} $, and any initial segment $ x = w \upharpoonright_{ \left| x \right| }  $ of an algorithmically random infinite string $ w $, there are sufficiently long middle segments $ y $ 
such that each $ y $ is unpredictable from the finite string $ x $ by a machine $ \mathbf{U} $ and a formal theory $ \mathbf{F} $,
where $ xy = w \upharpoonright_{ \left| x \right| + \left| y \right| } $.
Also note that, from the basic properties of $ \mathbf{K}\left( \cdot \right) $ in AIT, it is easy to show that if $ y $ is unpredictable from the finite string $ x $ by a machine $ \mathbf{U} $ and a formal theory $ \mathbf{F} $, then $ y $ is also \emph{uncomputable} from the finite string $ x $ by a machine $ \mathbf{U} $ and a formal theory $ \mathbf{F} $,
as long as $ \mathbf{K}\left( y \right) $ is sufficiently larger than the constant $ C $.

In this manner, we can formalise the unpredictability of models as:

\begin{definition}[Unpredictable model]\label{defUnpredictablemodels}
	Let $ \mathrm{P} $ be a learning algorithm that satisfies Definition~\ref{defLearningalgorithm} with the formal theory $ \mathbf{F} $.
	We say a model $ \mathrm{M} $ is \emph{unpredictable} from the learning algorithm $ \mathrm{P} $, available dataset $ \mathrm{D}_a $, and another model $ \mathrm{M}' $ by a machine $ \mathbf{U} $ and a formal theory $ \mathbf{F} $
	iff
	there is a fixed constant $ C < \mathbf{K}\left(  \mathrm{M} \right) $ (which only depends on the chosen machine and formal theory, but not on $ \mathrm{M} $, $ \mathrm{D}_a $, $ \mathrm{P} $, or $ \mathrm{M}' $) such that $ \mathbf{K}\left( \mathrm{M} \, \middle\vert \, \left< \mathrm{D}_a , \mathrm{P} , \mathrm{M}'  \right> \right) 
	\geq 
	\mathbf{K}\left(  \mathrm{M} \right) - C $.
\end{definition}

For example, suppose now that a model $ \mathrm{M}' $ is considered to be optimal for the available dataset $ \mathrm{D}_a $ according to the learning algorithm $ \mathrm{P} $ as in Definition~\ref{defOptimalmodelfromavailabledata}. 
One can easily demonstrate that there are infinitely many optimal models $ \mathrm{M} $ that are \emph{unpredictable} from the learning algorithm $ \mathrm{P} $, available dataset $ \mathrm{D}_a $, and model $ \mathrm{M}' $ by machine $ \mathbf{U} $ and formal theory $ \mathbf{F} $.
To this end, let $ \mathrm{P} $, $ \mathrm{D}_a $, $ \mathrm{M}' $, $ \mathbf{U} $, and $ \mathbf{F} $ be fixed.
Then, arbitrarily choose any sufficiently large constant $ C $ such that
$ \mathbf{K}\left( \left< \mathrm{D}_a , \mathrm{P} , \mathrm{M}'  \right> \right) \leq C - \mathbf{O}(1) $.
Remember the non-triviality condition in Section~\ref{sectionPerformancemeasure} for the performance measure $ f_{per} $.
Thus, from this assumption, there are infinitely many datasets for which the model $ \mathrm{M}' $ is not considered to be optimal.
Since we also assumed the condition in Sections~\ref{sectionDatageneratingprocess} and \ref{sectionPerformancemeasure} that the set $ \left\{ \mathrm{D}_{new} \right\} $ of all possible fresh data must be infinite computably enumerable given a fixed $ \mathrm{D}_a $, then there will be infinitely many optimal models $ \mathrm{M} $ of each respective $  \mathrm{D}_{total} $ for which $ \mathrm{M}' $ is not considered to be optimal.
Finally, just choose among these models $ \mathrm{M} $ those for which 
$ C + \mathbf{O}(1) \leq \mathbf{K}\left( \mathrm{M} \right) $, which can always be done because there can exist only a finite number of distinct models $ \mathrm{M}'' $ such that $ \mathbf{K}\left( \mathrm{M''} \right) \leq C + \mathbf{O}(1)  $ holds.

The existence of such unpredictable models $ \mathrm{M} $ for which the former model $ \mathrm{M}' $ is not optimal leads to the existence of a ``deceiving phenomenon'' that can be brought into play by the available dataset for which $ \mathrm{M}' $ is an optimal model.
The main idea of
a dataset $ \mathrm{D}_a $ being a \emph{deceiver} for a learning algorithm $ \mathrm{P} $ is that the prediction error on fresh data (error which is calculated from $ \mathrm{M}_{ \left( \mathrm{P} , \mathrm{D}_a \right) } $ on $ \mathrm{D}_{total} $ by the performance measure)
%in comparison to the actual optimal model $ \mathrm{M}_{ \left( \mathrm{P} , \mathrm{D}_{total} \right) } $ on $ \mathrm{D}_{total} $) 
becomes greater than $ \delta $,
where 
$ \mathrm{D}_{total} = \mathrm{D}_a \cup \mathrm{D}_{new} $,
$ \mathrm{D}_{new} $ is the dataset composed of the fresh data that $ \mathrm{P} $ has no access to in the learning process,
$ \delta \in Z_{ per }  $ is an arbitrary constant shown to exist at the end of Section~\ref{sectionPerformancemeasure}.
%and $ \mathrm{M}_{ \left( \mathrm{P}' , \mathrm{D} \right) } $ denotes the optimal model calculated by an arbitrary learning algorithm $ \mathrm{P}' $ from an arbitrary available dataset $ \mathrm{D} $.
In other words, a deceiving available dataset is one from which the learning algorithm (with the formal theory $ \mathbf{F} $ already included) could \emph{not} find an optimal model that performs optimally on fresh data, although it did find one that performs optimally on the available dataset.
This is formalised by:

\begin{definition}[Deceiving dataset]\label{defDeceivingdataset}
%	Let $ \mathrm{P} $ be an learning algorithm that satisfies Definition~\ref{defLearningalgorithm} with
%	a data generating source $ \mathcal{X}_{ \mathrm{D} } $,
%	a performance measure $ f_{per} $,
%	and a formal theory $ \mathbf{F} $.
	Let $ \mathcal{L}\left( \mathcal{X}_{ \mathrm{D} } , \mathcal{X}^m , f_{per} , \mathbf{F} , \mathrm{P} \right) $
	be a \emph{learning process} of the learning algorithm $ \mathrm{P} $ from the data generating source  $ \mathcal{X}_{ \mathrm{D} } $ as in Definition~\ref{defLearningprocess}.
	We say an available dataset $ \mathrm{D}_a $ is a \emph{deceiver} in the learning process $ \mathcal{L}\left( \mathcal{X}_{ \mathrm{D} } , \mathcal{X}^m , f_{per} , \mathbf{F} , \mathrm{P} \right) $
	iff:
	\begin{itemize}
		\item $ \mathrm{M}_{ \left( \mathrm{P} , \mathrm{D}_a \right) } $ is an \emph{optimal model} from the available dataset $ \mathrm{D}_a $ according to learning algorithm $ \mathrm{P} $ as in Definition~\ref{defOptimalmodelfromavailabledata};
		
%		\item and $  \mathbf{U}\left( \left< \mathbf{F} , \mathrm{M}_{ \left( \mathrm{P} , \mathrm{D}_a \right) } , \mathrm{D}_{total}  , p_{opt} \right> \right) = 0  $.
		
		\item and $ \mathrm{M}_{ \left( \mathrm{P} , \mathrm{D}_a \right) } $ is \emph{not} a \emph{globally optimal model} of the data generating source $ \mathcal{X}_{ \mathrm{D} } $ according to the performance measure $ f_{per} $ and formal theory $ \mathbf{F} $ as in Definition~\ref{defGlobaloptima}.
	\end{itemize}
\end{definition}

%\color{blue}
%Now, we can introduce the concept of deceiving datasets:

%\begin{definition}[informal]
%	We say that a dataset $ \mathrm{D}_a $ is a \emph{deceiver} for a learning algorithm $ \mathrm{P} $ if the prediction error on fresh data (calculated from $ \mathrm{M}_{ \left( \mathrm{P} , \mathrm{D}_a \right) } $ on $ \mathrm{D}_{total} $ in comparison to the actual optimal model $ \mathrm{M}_{ \left( \mathrm{P} , \mathrm{D}_{total} \right) } $ on $ \mathrm{D}_{total} $) is greater than $ \delta $,
%	where 
%	$ \mathrm{D}_{total} = \mathrm{D}_a \cup \mathrm{D}_{new} $,
%	$ \mathrm{D}_{new} $ is the dataset composed of the fresh data that $ \mathrm{P} $ has no access to in the learning process,
%	and $ \delta > 0  $ is a rational number such that $ \delta $ is less than or equal to the maximum generalisation/prediction error with respect to $ \mathrm{M}_{ \left( \mathrm{P} , \mathrm{D}'_a \right) } $, for any $ \mathrm{D}'_a $.
%	In addition, we say a dataset $ \mathrm{D}_a $ is an \emph{unpredictable deceiver} for a learning algorithm $ \mathrm{P} $ if it is a deceiver and the globally optimal model $ \mathrm{M}_{ \left( \mathrm{P} , \mathrm{D}_{total} \right) } $ is uncomputable from $ \mathrm{P} $, $ \delta  $, and $ \mathrm{M}_{ \left( \mathrm{P} , \mathrm{D}_a \right) } $ combined, given an arbitrarily chosen formal theory.\todo{Sentence edited.}
%\end{definition}

%\color{black}

Then, by combining both definitions, we obtain:

\begin{definition}[Unpredictably deceiving dataset]\label{defUnpredictabledeceiver}
	Let $ \mathcal{L}\left( \mathcal{X}_{ \mathrm{D} } , \mathcal{X}^m , f_{per} , \mathbf{F} , \mathrm{P} \right) $
	be a \emph{learning process} of the learning algorithm $ \mathrm{P} $ from the data generating source  $ \mathcal{X}_{ \mathrm{D} } $ as in Definition~\ref{defLearningprocess}.
	We say an available dataset $ \mathrm{D}_a $ is a \emph{unpredictable deceiver} in the learning process $ \mathcal{L}\left( \mathcal{X}_{ \mathrm{D} } , \mathcal{X}^m , f_{per} , \mathbf{F} , \mathrm{P} \right) $ for machine $ \mathbf{U} $
	iff:
	\begin{itemize}
		\item $ \mathrm{D}_a $ is a \emph{deceiver} in the learning process $ \mathcal{L}\left( \mathcal{X}_{ \mathrm{D} } , \mathcal{X}^m , f_{per} , \mathbf{F} , \mathrm{P} \right) $ as in Definition~\ref{defDeceivingdataset};
		
		\item and $ \mathrm{M}_{ \left( \mathrm{P} , \mathrm{D}_{total} \right) } $ is \emph{unpredictable} from the learning algorithm $ \mathrm{P} $, available dataset $ \mathrm{D}_a $, and model $ \mathrm{M}_{ \left( \mathrm{P} , \mathrm{D}_a \right) } $ by a machine $ \mathbf{U} $ and a formal theory $ \mathbf{F} $ as in Definition~\ref{defUnpredictablemodels}.
	\end{itemize}
\end{definition}

It is important to remark that, although demonstrating the existence of unpredictable deceivers in the case of an arbitrary data generating source $ \mathcal{X}_{ \mathrm{D} } $ (as in Definition~\ref{defDatageneratingsource}) is straightforward, it does not assert about how likely unpredictable deceivers occur.
In order to do so, one would need to assume conditions or constraints about the data generating source $ \mathcal{X}_{ \mathrm{D} } $ that ensures whether the probability distribution is more biased or more uniform toward unpredictably deceiving datasets.
Additionally, we have \emph{not} demonstrated that \emph{no} learning algorithm can make such a deceiving phenomenon unlikely to happen for certain specific data generating sources. 
It may be the case that a data generating source that is more biased toward non-deceivers enables one to construct a learning algorithm that can make the probability of occurring an unpredictable deceiver as low as one wishes.
This resembles the problem of the no free lunch theorem in data science and machine learning that we have discussed in Section~\ref{sectionMachinelearninginlargedatasets}.
For most practical purposes, some method, heuristics, or pre-processing is employed in order to e.g. reduce the scope of the problem, reduce the size of the datasets, select only a few viable candidates from the datasets or assume a priori constraints on the possible models.
All of these would imply some kind of bias that, by construction, constraints the overall complexity of the models, which in turn would avoid the necessary conditions for the no free lunch theorem to apply.

By following this scope of the problem, the challenge that we tackle in the next section is to study the probability of unpredictable deceivers for arbitrary learning algorithms on data generating sources that are biased toward low algorithmic complexity, as in Definition~\ref{defUniversaldatageneratingprocess}.
Besides assuring a formal version of the bias toward simplicity discussed in the previous paragraph that is also universal for the set of all computably constructible objects as shown by the algorithmic coding theorem, this kind of data generating source were already shown to overcome overfitting for available datasets (see also Section~\ref{sectionMachinelearningandalgorithmicinformation}).
However, we shall demonstrate in the following section that no learning algorithm can make the probability of unpredictably deceiving datasets as negligible as one wishes for data generating sources satisfying Definition~\ref{defUniversaldatageneratingprocess}.

\section{A deceiving dataset for learning algorithms}\label{sectionResults}

For the sake of simplifying our forthcoming proofs, we employ a slight variation of the Busy Beaver function found in \cite{Chaitin2012,Abrahao2017publishednat,Bennett1988}: let $ BB : \, \mathbb{N} \to \, \mathbb{N} $ be a Busy Beaver function that calculates the largest integer that a program $ p $ with length $ \leq n \in \mathbb{N} $ can output running on universal Turing machine $ \mathbf{U} $ and adds $ 1 $ to this largest value \cite{Abrahao2016nat}. 
That is, $ BB\left( n \right) = 1 + \max \left\{ \mathbf{U}\left( p \right) \, \middle\vert \, n \geq \left| p \right| \right\} $.

In order to achieve the following proofs, it is important to remember the usual properties of such a function $ BB\left( \cdot \right) $  and the halting probability (or Chaitin's Omega number) 
$ \Omega = \sum_{ \mathbf{U}\left( p \right) \downarrow }  2^{ - \left| p \right| } $ \cite{Calude2002,Chaitin2004,Chaitin1975}.
This function inherits the same properties described in \cite[Section~4.5]{Abrahao2017publishednat}, \cite[Lemma~5.1]{Abrahao2017publishednat}, and \cite{Abrahao2016nat}.
In particular, we know from the construction of function $ BB\left( n \right) = 1 + \max \left\{ \mathbf{U}\left( p \right) \, \middle\vert \, n \geq \left| p \right| \right\} $ and  the literature of AIT and computability theory that the following properties hold:

\begin{lemma}\label{lemmaBB1}
	For every $ n , \, m \in \mathbb{N} $ with $  BB\left( n \right) \leq m $ and $ n \geq k $, we have that
	\begin{equation}
		BB\left( n \right) > BB\left( k \right)
		\text{ ,}
	\end{equation}
	\begin{equation}
		\mathbf{K}\left( BB\left( n \right) \right) > n
	\end{equation}
	and 
	\begin{equation}
		\mathbf{K}\left( m \right) > n
	\end{equation}
	hold.
\end{lemma}

\begin{lemma}\label{lemmaBB2}
	There are programs $ p_{ BB } $ and $ p_{ halt } $ such that for every $ x \in \mathbb{N} $: 
	\begin{itemize}
		\item $ \mathbf{U}\left(  \left< \left( \sum_{ \mathbf{U}\left( p \right) \downarrow }  2^{ - \left| p \right| } \right) \upharpoonright_x , \, p_{ BB } \right> \right) 
		= 
		BB\left( x \right) $ and;
		
		\item $ p_{ halt } $ decides whether or not a program $ p $ halts program $ w $, where $ \left| w \right| \leq x $, i.e., for every program $ w $, \\
		\begin{equation}
			\begin{aligned}
				\mathbf{U}\left(  \left< \left( \sum_{ \mathbf{U}\left( p \right) \downarrow }  2^{ - \left| p \right| } \right) \upharpoonright_x , \, w \upharpoonright_{ x } , \, p_{ halt } \right> \right) 
				= 
				\begin{cases}
				1 
				& \textnormal{, if $ \mathbf{U}\left( w \upharpoonright_{ x } \right) \downarrow $ } \\
				0
				& \textnormal{, if $ \mathbf{U}\left( w \upharpoonright_{ x } \right) \uparrow $ }
				\end{cases}
%				\text{ .}
			\end{aligned}
		\end{equation}
		\noindent hold.
	\end{itemize}
\end{lemma}

The next Lemma~\ref{lemmaD_a} shows that
there is an algorithm that can build an available dataset $ \mathrm{D}_a $ with arbitrary $ n $ such that $ \left| \mathrm{D}_a \right| $ is greater than any value any program of length $ \leq n $ can calculate and the algorithmic complexity of the optimal model $ \mathrm{M}_{ \left( \mathrm{P} , \mathrm{D}_a \right) } $ returned by $ \mathrm{P} $ from $ \mathrm{D}_a $ is upper bounded by an arbitrary value $ k \leq n $. 
%where the value\todo{update} of $ k $ is considered to be small by the learning algorithm so that $ \mathrm{M}_{ \left( \mathrm{P} , \mathrm{D}_a \right) } $ would avoid overfitting \cite{HernandezOrozco2021}.
In other words, even if the underlying data generating process \cite{Goodfellow2016} of the available dataset is more complex than the learning algorithm itself, the algorithmic information necessary to allow this process to generate a dataset whose model calculated by the learning algorithm is optimal is not greater than $ k \leq n $.

\begin{lemma}\label{lemmaD_a}
	Let $ \mathrm{P} $ be an arbitrary learning algorithm that satisfies Definition~\ref{defLearningalgorithm} such that $ \mathcal{X}^m $ is computable (or, equivalently without loss of generality, $ \mathcal{X}^m = \emptyset $).
	Let $ n \in \mathbb{N} $ be an arbitrarily large constant
	such that
	$ n \geq \min\left\{ \mathbf{K}\left( \mathrm{M} \right) \, \middle\vert \,  \mathrm{M} \text{ satisfies Definition~\ref{defModel}} \right\} $ holds. 
%	Let $ n \in \mathbb{N} $ be an arbitrary constant such that $  f\left( n \right) \leq n $, where $ f : \mathbb{N} \to \mathbb{N} $ is a total computable function already included in $ \mathrm{P} $.
	Then, there is an available dataset $ \mathrm{D}_a $ such that 
	\begin{equation}\label{equationlemmaD_a1}
		\left| \mathrm{D}_a \right| \geq BB\left( n \right)\text{ ,}
	\end{equation}
	\begin{equation}\label{equationlemmaD_a2}
		\mathbf{K}\left( \mathrm{D}_a \right)
		\leq
		\mathbf{K}\left( \mathrm{P} \right) + n + \mathbf{K}\left( n \right) + \mathbf{O}\left( 1 \right)
		\leq
		\mathbf{K}\left( \mathrm{P} \right) + n + \mathbf{O}\left( \log\left( n \right) \right)
		\text{ ,}
	\end{equation}
	and 
%	$
	\begin{equation}\label{equationlemmaD_a3}
		 \begin{aligned}
			 \mathbf{K}\left( \mathrm{M}_{ \left( \mathrm{P} , \mathrm{D}_a \right) } \right) 
			 \leq
			 n
%			 f\left( n \right)
%			 \leq \\
%			 \mathbf{K}\left( \mathrm{P} \right)
%			 +
%			 \mathbf{K}\left( \mathrm{D}_a \right)
%			 +
%			 \mathbf{O}\left( 1 \right)
%			 \leq \\
%			 2 \, \mathbf{K}\left( \mathrm{P} \right)
%			 +
%			 n
%			 +
%			 \mathbf{O}\left( 1 \right)
			 \text{ ,}
		 \end{aligned}
	\end{equation}
%	$ 
	hold,
	where
	the optimal model $ \mathrm{M}_{ \left( \mathrm{P} , \mathrm{D}_a \right) } $ satisfies Definition~\ref{defOptimalmodelfromavailabledata}. 
	
	\begin{proof}
		Let $ h $ be a bit string that represents an algorithm running on a prefix universal Turing machine $ \mathbf{U} $ that receives $ \mathrm{P} $ and $ \Omega \upharpoonright_n $ as inputs, 
		where $ \Omega $ is the halting probability
		and 
		$ \mathrm{P} $ satisfies Definition~\ref{defLearningalgorithm}.
%		and $ x \upharpoonright_n $ denotes the sequence of length $n$ corresponding to the first $ n $ fractional bits of the real number $ x $.\todo{notation might be redundant.}
		%		and 
		%		$ \left| m - 2 \, \big( \mathbf{K}\left( \mathrm{P} \right) + \mathbf{K}\left( \mathrm{D}_a \right) + \mathbf{K}\left(  \delta \right)  + \mathbf{O}\left( 1 \right) \big) + c''  \right| = \mathbf{O}\left( 1 \right) $.
		Next, $ h $ computably enumerates the halting programs until it finds the first $ n' \geq n  $ such that 
		\begin{equation}
			\begin{aligned}
				\Omega \upharpoonright_n
				\leq 
				\left( \underset{ \scriptsize
					\begin{array}{c}
					\mathbf{U}\left( p \right) \downarrow \\
					p \in {\text{Halt}_h}_i 
					\end{array}
				}{ \sum }  2^{ - \left| p \right| } \right) \upharpoonright_{n'}
				\text{ ,}
			\end{aligned}
		\end{equation}
		where $ {\text{Halt}_h}_i $ is the set of halting programs that $ h $ is enumerating at each $ i $-th iteration.
		In the next step, $ h $ starts a dovetailing enumeration of: 
		all $ \mathrm{D}_a $ such that  $ \left| \mathrm{D}_a \right| \geq BB\left( n \right) $;
		and their respective optimal models $ \mathrm{M}_{ \left( \mathrm{P} , \mathrm{D}_a \right) } $ calculated by $ \mathrm{P} $ as in Definitions~\ref{defOptimalmodelfromavailabledata} and~\ref{defLearningalgorithm}
		such that $ \mathbf{K}\left( \mathrm{M}_{ \left( \mathrm{P} , \mathrm{D}_a \right) } \right) \leq n $ holds.
		Finally, $ h $ returns the first $ \mathrm{D}_a $ that satisfies these two criteria.
		From Lemma~\ref{lemmaBB2}, we know that program $ h $ can always calculate $ BB\left( n \right) $ and $ \mathbf{K}\left( \mathrm{M}_{ \left( \mathrm{P} , \mathrm{D}_a \right) } \right) $, where $ \left| \mathrm{M}_{ \left( \mathrm{P} , \mathrm{D}_a \right) } \right| \leq n $.
		From our assumption on $ n $ and from the extensibility condition of the performance measure in Section~\ref{sectionPerformancemeasure}, we know there are an infinite computably enumerable set of available datasets for which $ \left| \mathrm{D}_a \right| \geq BB\left( n \right) $ and $ \mathbf{K}\left( \mathrm{M}_{ \left( \mathrm{P} , \mathrm{D}_a \right) } \right) \leq n $ hold at the same time.
%		Now, remember that, for every $ x \in \mathbb{N} $ and program $ p $ with $ \left| p \right| \leq x $, $ \left( \sum_{ \mathbf{U}\left( p \right) \downarrow }  2^{ - \left| p \right| } \right) \upharpoonright_x $ is sufficient for computing $ BB\left( x \right) $ and deciding whether or not a program $ p $ halts \todo{Sentence edited}.
		Therefore, program $ h $ is well defined (i.e., it always halts).
		Equation~\ref{equationlemmaD_a2} follows straightforwardly from our construction of $ h $ and basic properties of $ \mathbf{K}\left( \cdot \right) $.
	\end{proof}
\end{lemma}

The last step before reaching the central Theorems~\ref{thmDeceiverDatasets} and \ref{thmDeceivingprobability} is to show in Lemma~\ref{lemmaD_total} below that for every computable learning process, there is a sufficiently large dataset $ \mathrm{D}_{ total } $ whose underlying data generating process can be as complex as one wishes in comparison to the learning algorithm $ \mathrm{P} $.
%If the learning process is computable, then the constant $ C_{ \mathcal{X} } $ assumes the value $ 0 $.
In addition, even though the available dataset $ \mathrm{D}_a \subset \mathrm{D}_{ total } $ might suffice to allow $ \mathrm{P} $ to estimate an optimal model $ \mathrm{M}_{ \left( \mathrm{P} , \mathrm{D}_a \right) } $, fresh data in $ \mathrm{D}_{new} \subset \mathrm{D}_{ total } $ can render the algorithmic complexity of the actual globally optimal model $ \mathrm{M}_{ \left( \mathrm{P} , \mathrm{D}_{ total } \right) } $ as large as one wishes.

\begin{lemma}\label{lemmaD_total}
	Let $ \mathrm{P} $ be an arbitrary learning algorithm that satisfies Definition~\ref{defLearningalgorithm} such that $ \mathcal{X}^m $ is computable (or, equivalently without loss of generality, $ \mathcal{X}^m = \emptyset $).
%	Let $ \mathcal{L}\left( \mathcal{X}_{ \mathrm{D} } , \mathcal{X}^m , f_{per} , \mathbf{F} , \mathrm{P} \right) $
%	be a computable learning process of the learning algorithm $ \mathrm{P} $ from the data generating source  $ \mathcal{X}_{ \mathrm{D} } $ as in Definition~\ref{defComputablelearningprocess}.
%	Let $ \mathrm{D}_a $ be an arbitrary dataset that is available to the learning algorithm $ \mathrm{P} $, allowing it to calculate the optimal model $ \mathrm{M}_{ \left( \mathrm{P} , \mathrm{D}_a \right) } $.
%	Let $ \delta > 0  $ be a rational number such that $ \delta $ is less than or equal to the maximum generalisation/prediction error with respect to $ \mathrm{M}_{ \left( \mathrm{P} , \mathrm{D}'_a \right) } $, for any $ \mathrm{D}'_a $.\todo{update}
	Let $ m \in \mathbb{N} $ be an arbitrary constant.
	Then, there is a dataset $ \mathrm{D}_{ total } = \left( \mathrm{D}_a \, , \, \mathrm{D}_{new} \right) $ such that 
	the available dataset $ \mathrm{D}_a $ is a deceiver according to Definition~\ref{defDeceivingdataset}
%	the performance measure (or prediction error) of the actual globally optimal model $ \mathrm{M}_{ \left( \mathrm{P} , \mathrm{D}_{total} \right) } $ with respect to $ \mathrm{M}_{ \left( \mathrm{P} , \mathrm{D}_a \right) } $ is greater than or equal to $ \delta $\todo{update} 
	and the inequality
	\begin{equation}\label{equationlemmaD_total} 
	\begin{aligned}
		m
		-
		\mathbf{K}\left( \mathrm{P} \right) 
		- 
		\mathbf{K}\left( \mathrm{D}_a \right) 
%		- 
%		\mathbf{K}\left(  \delta \right)  
		- 
		\mathbf{O}\left( 1 \right)
		< \\
		\mathbf{K}\left(  \mathrm{M}_{ \left( \mathrm{P} , \mathrm{D}_{ total } \right) }  \right) 
		\leq \\
		\mathbf{K}\left(  \mathrm{D}_{ total }  \right) 
		+ 
		\mathbf{K}\left( \mathrm{P} \right) 
		+ 
		\mathbf{O}\left( 1 \right) 
		\leq  \\
		m
		+ \mathbf{O}\left( \log\left( m \right) \right)
		+
		2 \, \mathbf{K}\left( \mathrm{P} \right) 
%		+
%		\mathbf{K}\left( \delta \right)
		+
		\mathbf{K}\left(  \mathrm{D}_a  \right) 
		+ 
		\mathbf{O}\left( 1 \right)
%		4 \, \mathbf{K}\left( \mathrm{P} \right) + 3 \, \mathbf{K}\left( \mathrm{D}_a \right) + 3 \, \mathbf{K}\left(  \delta \right) + \mathbf{O}\left( 1 \right)
%		+ 
%		2 \, c''
%		+\\
%		\mathbf{O}\left( \log_2\left( \mathbf{K}\left( \mathrm{P} \right) + \mathbf{K}\left( \mathrm{D}_a \right) + \mathbf{K}\left(  \delta \right)  + \mathbf{O}\left( 1 \right) + c'' \right) \right)
%		+
%		\mathbf{O}\left( 1 \right)
%		\text{ ,}
	\end{aligned}
	\end{equation}
%	and
%	\begin{equation}
%		\mathbf{K}\left(  \mathrm{D}_{ total }  \right)
%		\leq
%		\log_2\left(  \left| \mathrm{D}_{ a } \right| \right)
%		\text{ ,}
%	\end{equation}
	holds, where 
%	$ \mathrm{D}_{new} $ is the dataset composed of the fresh data that $ \mathrm{P} $ has no access to in the learning process, and 
	$ \mathrm{M}_{ \left( \mathrm{P} , \mathrm{D}_{total} \right) } $ is a globally optimal model that satisfies Definition~\ref{defGlobaloptima} 
%	with $ \mathcal{L}\left( \mathcal{X}_{ \mathrm{D} } , \mathcal{X}^m , f_{per} , \mathbf{F} , \mathrm{P} \right) $ and $ \mathrm{D}_{ total } $.
	
	\begin{proof}
		Let $ h' $ be a bit string that represents an algorithm running on a prefix universal Turing machine $ \mathbf{U} $ that receives $ \mathrm{P} $, $ \mathrm{D}_a $, and $ \Omega \upharpoonright_m $ as inputs.
%		where $ \Omega $ is the halting probability (or Chaitin's Omega number) \cite{Calude2002,Chaitin2004,Chaitin1975}. 
%		and 
%		$ \left| m - 2 \, \big( \mathbf{K}\left( \mathrm{P} \right) + \mathbf{K}\left( \mathrm{D}_a \right) + \mathbf{K}\left(  \delta \right)  + \mathbf{O}\left( 1 \right) \big) + c''  \right| = \mathbf{O}\left( 1 \right) $.
		In the next step, $ h' $ starts a dovetailing enumeration of all $  \mathrm{D}_{ total } = \left( \mathrm{D}_a \, , \, \mathrm{D}_{new} \right) $ and their respective optimal models $ \mathrm{M}_{ \left( \mathrm{P} , \mathrm{D}_{total} \right) } $.
%		such that $ \mathrm{M}_{ \left( \mathrm{P} , \mathrm{D}_a \right) } $ is \emph{not} a globally optimal model according Definition~\ref{defGlobaloptima}, although $ \mathrm{M}_{ \left( \mathrm{P} , \mathrm{D}_a \right) } $ is an optimal model according to Definition~\ref{defOptimalmodelfromavailabledata}.
%		the prediction error with respect to $ \mathrm{M}_{ \left( \mathrm{P} , \mathrm{D}_a \right) } $ is larger than or equal to $ \delta $.
		Next, $ h' $ computably enumerates the halting programs until it finds the first $ m' \geq m $ such that 
		\begin{equation}
			\begin{aligned}
				\Omega \upharpoonright_m
				\leq 
				\left( \underset{ \scriptsize
					\begin{array}{c}
					\mathbf{U}\left( p \right) \downarrow \\
					p \in {\text{Halt}_{ h' }}_i
					\end{array}
				}{ \sum }  2^{ - \left| p \right| } \right) \upharpoonright_{m'}
				\text{ ,}
			\end{aligned}
		\end{equation}
		where $ {\text{Halt}_{ h' }}_i $ is the set of halting programs that $ h' $ is enumerating at each $ i $-th iteration.
		Then, $ h' $ stops this dovetailing enumeration when it finds the $ BB\left( m \right) $-th distinct optimal model $ \mathrm{M}_{ \left( \mathrm{P} , \mathrm{D}_{total} \right) } $
		for which the available dataset $ \mathrm{D}_a $ is a deceiver according to Definition~\ref{defDeceivingdataset}.
%		such that $ \mathrm{M}_{ \left( \mathrm{P} , \mathrm{D}_a \right) } $ is \emph{not} a globally optimal model according Definition~\ref{defGlobaloptima}, although $ \mathrm{M}_{ \left( \mathrm{P} , \mathrm{D}_a \right) } $ is an optimal model according to Definition~\ref{defOptimalmodelfromavailabledata}.
%		the prediction error with respect to $ \mathrm{M}_{ \left( \mathrm{P} , \mathrm{D}_a \right) } $ is greater than or equal to $ \delta $, 
		Finally, $ h' $ returns as output the dataset $ \mathrm{D}_{total} $ relative to this $ BB\left( m \right) $-th globally optimal model.
		Note that $ h' $ can always calculate $ BB\left( m \right) $ because of Lemma~\ref{lemmaBB2}.
		From the non-triviality condition of the performance measure in Section~\ref{sectionPerformancemeasure}, we know there is an infinite computably enumerable set of datasets $ \mathrm{D}_{total} $ for which $ \mathrm{D}_a $ is a deceiver according to Definition~\ref{defDeceivingdataset}.
		Therefore, program $ h' $ is well defined.
		Now, let $ h'' $ be a program that receives $ \mathrm{P} $, $ \mathrm{D}_a $, and $  \mathrm{M}_{ \left( \mathrm{P} , \mathrm{D}_{ total } \right) } $ as inputs.
		Then, by using the above program $ h' $ as its subroutine,  $ h'' $ starts a dovetailing enumeration of all $  \mathrm{D}'_{ total } = \left( \mathrm{D}'_a \, , \, \mathrm{D}'_{new} \right) $, their respective optimal models $ \mathrm{M}'_{ \left( \mathrm{P} , \mathrm{D}'_{total} \right) } $, and selects only those
		for which the available dataset $ \mathrm{D}'_a $ is a deceiver according to Definition~\ref{defDeceivingdataset}.
		$ h'' $ stops this enumeration when it finds the first of these distinct optimal models  $ \mathrm{M}'_{ \left( \mathrm{P} , \mathrm{D}_{total} \right) } $ for which the available dataset $ \mathrm{D}_a $ is a deceiver such that $ \mathrm{M}'_{ \left( \mathrm{P} , \mathrm{D}_{total} \right) }  = \mathrm{M}_{ \left( \mathrm{P} , \mathrm{D}_{ total } \right) } $.
		Let this first optimal model  $ \mathrm{M}'_{ \left( \mathrm{P} , \mathrm{D}_{total} \right) } $ be the $ k $-th one in the previous enumeration.
		Finally, $ h'' $ returns $ k $ as output.
		Note that if $ \mathrm{M}_{ \left( \mathrm{P} , \mathrm{D}_{ total } \right) } $ is the $ BB\left( m \right) $-th globally optimal model calculated by program $ h' $, then by construction $ h'' $ returns $ BB\left( m \right) $ as output.
%		$ h'' $ computably enumerates the halting programs constructing left approximations  (i.e., lower bounds) 
%		\begin{equation}
%			 \Omega_i = \underset{ \scriptsize
%				\begin{array}{c}
%				\mathbf{U}\left( p \right) \downarrow \\
%				p \in {\text{Halt}_{ h'' }}_i 
%				\end{array}
%			}{ \sum }  2^{ - \left| p \right| } 
%		\end{equation}
%		for  $ \Omega $, and for each $ i $-th iteration of this enumeration
%		it runs program
		Therefore, we will have it that the inequality $
%		\begin{equation}
		m
		-
		\mathbf{K}\left( \mathrm{P} \right) 
		- 
		\mathbf{K}\left( \mathrm{D}_a \right) 
%		- 
%		\mathbf{K}\left(  \delta \right)  
		- 
		\mathbf{O}\left( 1 \right)
		<
		\mathbf{K}\left(  \mathrm{M}_{ \left( \mathrm{P} , \mathrm{D}_{ total } \right) }  \right)
%		\end{equation}
		$ holds because of basic properties of $ \mathbf{K}\left( \cdot \right) $, our construction of the programs $ h' $ and $ h'' $, and Lemma~\ref{lemmaBB1}.
		The inequality $ \mathbf{K}\left(  \mathrm{M}_{ \left( \mathrm{P} , \mathrm{D}_{ total } \right) }  \right) 
		\leq 
		\mathbf{K}\left(  \mathrm{D}_{ total }  \right) 
		+ 
		\mathbf{K}\left( \mathrm{P} \right) 
		+ 
		\mathbf{O}\left( 1 \right) $
		follows from Notation~\ref{defModelcalculatedbylearningalgorithm} of $ \mathrm{M}_{ \left( \mathrm{P} , \mathrm{D}_{ total } \right) } $, Definition~\ref{defLearningalgorithm} of learning algorithm, and basic properties of $ \mathbf{K}\left( \cdot \right) $.
%		where $ \left< \cdot , \cdot \right> $ denotes an arbitrarily chosen pairing function.
		Finally, the inequality 
		$ 
		\mathbf{K}\left(  \mathrm{D}_{ total }  \right) 
		+ 
		\mathbf{K}\left( \mathrm{P} \right) 
		+ 
		\mathbf{O}\left( 1 \right)
		\leq
		m
		+ \mathbf{O}\left( \log\left( m \right) \right)
		+
		2 \, \mathbf{K}\left( \mathrm{P} \right) 
%		+
%		\mathbf{K}\left( \delta \right)
		+
		\mathbf{K}\left(  \mathrm{D}_a  \right) 
		+ 
		\mathbf{O}\left( 1 \right)
		$ 
		follows directly from the inequality
		$ 
		\mathbf{K}\left(  \mathrm{D}_{ total }  \right) 
%		+ 
%		\mathbf{K}\left( \mathrm{P} \right) 
%		+ 
%		\mathbf{O}\left( 1 \right)
		\leq
		m
		+ \mathbf{O}\left( \log\left( m \right) \right)
		+
		\mathbf{K}\left( \mathrm{P} \right) 
		%		+
		%		\mathbf{K}\left( \delta \right)
		+
		\mathbf{K}\left(  \mathrm{D}_a  \right) 
		+ 
		\mathbf{O}\left( 1 \right)
		$,
		which in turn follows straightforwardly 
		from our construction of the program $ h' $ and basic properties of $ \mathbf{K}\left( \cdot \right) $.
%		Finally, by replacing $m $ with its upper bound from our construction, we obtain the last remaining inequality from the inequality $ \mathbf{K}\left(  \mathrm{D}_{ total }  \right) 
%		+ 
%		\mathbf{K}\left( \mathrm{P} \right) 
%		+ 
%		\mathbf{O}\left( 1 \right) 
%		\leq  
%		2 \, \mathbf{K}\left( \mathrm{P} \right) + \mathbf{K}\left( \mathrm{D}_a \right) + \mathbf{K}\left(  \delta \right) + \mathbf{O}\left( 1 \right)
%		+ 
%		m
%		+
%		\mathbf{O}\left( \log_2\left( m \right) \right)
%		+
%		\mathbf{O}\left( 1 \right) $, which follows from our construction of the program $ h' $ so that $ \mathbf{U}\left( \left<  \mathrm{P} , \delta , \Omega \upharpoonright_m , \mathrm{D}_a , h' \right> \right) = \mathrm{D}_{ total } $ holds.
	\end{proof}
\end{lemma}

The main idea of following Theorem~\ref{thmDeceiverDatasets} is that even though the available dataset $ \mathrm{D}_a $ is constructed in such a way that it satisfies Lemma~\ref{lemmaD_a}, the fresh dataset $ \mathrm{D}_{new} $ is added to $ \mathrm{D}_a $ so that the algorithm that computes the underlying data generating process of $ \mathrm{D}_{ total } = \left( \mathrm{D}_a \, , \, \mathrm{D}_{new} \right) $ searches for the fresh dataset $ \mathrm{D}_{new} $ that satisfies Lemma~\ref{lemmaD_total}. 
%and from which the prediction error given by the learning algorithm is maximal.
Thus, the underlying data generating process that constructs $ \mathrm{D}_{total} $ deceives the learning algorithm $ \mathrm{P} $ into ``thinking'' that an optimal model can be found from $ \mathrm{D}_a $,
while in fact the actual globally optimal model is unpredictable. 
%and its value diverges from the locally optimal model $ \mathrm{M}_{ \left( \mathrm{P} , \mathrm{D}_a \right) } $ by an arbitrary value $ \delta $.

\begin{theorem}\label{thmDeceiverDatasets}
	Let $ \mathrm{P} $ be an arbitrary learning algorithm that satisfies Definition~\ref{defLearningalgorithm} such that $ \mathcal{X}^m $ is computable (or, equivalently without loss of generality, $ \mathcal{X}^m = \emptyset $).
%	Let $ \delta > 0  $ be a rational number such that $ \delta $ is less than or equal to the maximum generalisation/prediction error with respect to $ \mathrm{M}_{ \left( \mathrm{P} , \mathrm{D}_a \right) } $.
%	For every learning algorithm $ \mathrm{P} $, 
	Then, there is a sufficiently large dataset $ \mathrm{D}_{ total } = \left( \mathrm{D}_a \, , \, \mathrm{D}_{new} \right) $ satisfying Lemma~\ref{lemmaD_total} such that: 
	\begin{enumerate}
	\item $ \mathrm{P} $ calculates an optimal model $ \mathrm{M}_{ \left( \mathrm{P} , \mathrm{D}_a \right) } $ from the available dataset $ \mathrm{D}_a $, which satisfies Lemma~\ref{lemmaD_a}; 
	
%	\item the prediction error on fresh data (calculated from $ \mathrm{M}_{ \left( \mathrm{P} , \mathrm{D}_a \right) } $ on $ \mathrm{D}_{total} $ in comparison with the actual globally optimal model $ \mathrm{M}_{ \left( \mathrm{P} , \mathrm{D}_{total} \right) } $ on $ \mathrm{D}_{total} $) is greater than or equal to $ \delta $,
%	where $ \delta > 0  $ is a rational number such that $ \delta $ is less than or equal to the maximum generalisation/prediction error with respect to $ \mathrm{M}_{ \left( \mathrm{P} , \mathrm{D}'_a \right) } $, for any $ \mathrm{D}'_a $;
	
	\item the dataset $ \mathrm{D}_a $ is an unpredictable deceiver according to Definition~\ref{defUnpredictabledeceiver}. 
%	so that
%	the globally optimal model $ \mathrm{M}_{ \left( \mathrm{P} , \mathrm{D}_{total} \right) } $ is uncomputable from $ \mathrm{P} $, $ \delta  $, and $ \mathrm{M}_{ \left( \mathrm{P} , \mathrm{D}_a \right) } $ combined.

	\item and the inequality
	\begin{equation}\label{equationthmDeceiverDatasets3}
		\left| \mathrm{D}_a \right| 
		\geq 
		BB\big(
		\mathbf{K}\left( \mathrm{D}_{total} \right) 
		- 
		\mathbf{O}\left( \mathbf{K}\left( \mathrm{P} \right)
%		+
%		\mathbf{K}\left( \delta \right) 
		\right)
		\big)
	\end{equation} 
	holds;
	
	\end{enumerate}

	\begin{proof}
		Let $ q $ be a bit string that represents an algorithm running on a prefix universal Turing machine $ \mathbf{U} $ that receives $ \mathrm{P} $, $ \Omega \upharpoonright_n $, and $ \Omega \upharpoonright^m_n $ as inputs.
%		where $ c > m - n  > c_\Omega + c' $ such that $ \mathbf{K}\left( \Omega \upharpoonright_x \right) \geq x - c_\Omega $ holds for every $ x $ and
%		$ c $ is an arbitrary constant.
		Then, it runs the program $ h $ from the proof of Lemma~\ref{lemmaD_a} and calculates $ \mathbf{U}\left( \left< \mathrm{P} , \Omega_n , h \right> \right) = \mathrm{D}_a $.
		Finally, $ q $ returns $ \mathrm{D}_{ total } = \mathbf{U}\left( \left<  \mathrm{P} , \Omega \upharpoonright_m , \mathbf{U}\left( \left< \mathrm{P} , \Omega_n , h \right> \right) , h' \right> \right)  $, where $ h' $ is the program as in the proof of Lemma~\ref{lemmaD_total}.\footnote{ Note that $ \Omega \upharpoonright_m $ is the concatenation of $ \Omega \upharpoonright_n $ with $ \Omega \upharpoonright^m_n $.}
		From the proofs of Lemmas~\ref{lemmaD_a} and \ref{lemmaD_total}, one can straightforwardly demonstrate that program $ q $ is well defined. 
		Therefore, $ \mathrm{M}_{ \left( \mathrm{P} , \mathrm{D}_a \right) } $ is an optimal model from the available dataset $ \mathrm{D}_a $, which satisfies Lemma~\ref{lemmaD_a}.
		Now, let $ c' $ be arbitrarily large.
		Remember that $ n $ in Lemma~\ref{lemmaD_a} was arbitrary as long as it is sufficiently larger than a constant that only depends on the class of possible models for $ \mathrm{P} $ and machine $ \mathbf{U} $.
		Thus, let $ n $ be the minimum sufficiently large value in comparison to $ \mathbf{K}\left( \mathrm{P} \right) $ such that
		\begin{equation}\label{equationthm1Basicassumptiononn}
			\mathbf{O}\left( 
			\mathbf{K}\left( \mathrm{P} \right) 
	%		+ 
	%		\mathbf{K}\left(  \delta \right)  
			\right)
	%		\leq
	%		f\left( n \right)
			\geq n
			\geq
			\mathbf{K}\left( \mathrm{P} \right)
			\text{ ,}
		\end{equation}
		\begin{equation}
			\mathbf{K}\left( n \right) \leq n
			\text{ ,}
		\end{equation}
		and
		\begin{equation}
			n \geq \min\left\{ \mathbf{K}\left( \mathrm{M} \right) \, \middle\vert \,  \mathrm{M} \text{ satisfies Definition~\ref{defModel}} \right\}
		\end{equation}
		hold, 
		where 
%		$ f $ is a total computable function 
		Lemma~\ref{lemmaD_a} is satisfied for this  $ n \in \mathbb{N} $. 
		Let $ m $ be sufficiently greater than $ n $ such that there is $ c $
		with 
		\begin{equation}\label{equationInequalityforc}
			\begin{aligned}
				\mathbf{O}\left( \mathbf{K}\left( \mathrm{P} \right) \right)
				\geq \\
				c 
				\geq \\ 
				m - n - \mathbf{O}\left( \log\left( n \right) \right) > \\
				c_\Omega + c' > \\
				4 \, \mathbf{K}\left( \mathrm{P} \right) 
		%		+ \mathbf{K}\left( \delta \right) 
				+ 3 \, n
				+ \mathbf{O}\left( 1 \right)
				\text{ ,}
			\end{aligned}
		\end{equation} 
		where
%		$ \delta $ satisfies 
		Lemma~\ref{lemmaD_total} is satisfied for $ m $ and
%		$ n $ satisfies Lemma~\ref{lemmaD_a},  
		$ \mathbf{K}\left( \Omega \upharpoonright_x \right) \geq x - c_\Omega $ holds for every $ x $.
%		and
%		$ c $ is an arbitrary constant.
		Therefore, from basic properties of $ \mathbf{K}\left( \cdot \right) $, Equation~\ref{equationInequalityforc}, and our construction of program $ q $, we have it that
		\begin{equation}\label{equationLaststepthm1}
			\mathbf{K}\left( \mathrm{D}_{ total } \right)
			\leq
			\mathbf{K}\left( \mathrm{P} \right)
	%		+
	%		\mathbf{K}\left( \delta \right)
			+
			n
			+ \mathbf{O}\left( \log\left( n \right) \right)
	%		+
	%		\mathbf{K}\left( \mathrm{P} \right)
			+
			c
%			+ \mathbf{O}\left( \log\left( c \right) \right)
			+
			\mathbf{O}\left( 1 \right)
			\text{ .}
		\end{equation}
%		Respectively from Equations~\ref{equationlemmaD_a2} and~\ref{equationlemmaD_a3} in Lemma~\ref{lemmaD_a}, 
%		from Equation~\ref{equationInequalityforc},
%		from Equation~\ref{equationlemmaD_a2} in Lemma~\ref{lemmaD_a},
%		and from Equation~\ref{equationlemmaD_total} in Lemma~\ref{lemmaD_total}, 
		We also have it that
%		the prediction error calculated from $ \mathrm{M}_{ \left( \mathrm{P} , \mathrm{D}_a \right) } $ on $ \mathrm{D}_{total} $ in comparison with the actual optimal model $ \mathrm{M}_{ \left( \mathrm{P} , \mathrm{D}_{total} \right) } $ on $ \mathrm{D}_{total} $ is larger than or equal to $ \delta $ such that
		\begin{equation}
			\begin{minipage}{0.9\linewidth}
				\begin{align}
					\mathbf{K}\left( \mathrm{P} \right) 
					+ \mathbf{K}\left( \mathrm{D}_a \right)
					+ \mathbf{K}\left(  \mathrm{M}_{ \left( \mathrm{P} , \mathrm{D}_a \right) }  \right)
					+ \mathbf{O}\left( 1 \right)
					\leq \tag{\ref{equationthmDeceiverDatasets}.1}\label{equationthmDeceiverDatasets.1}\\
					2 \, \mathbf{K}\left( \mathrm{P} \right) 
					+ 3 \, n
					+ \mathbf{O}\left( 1 \right)
					< \tag{\ref{equationthmDeceiverDatasets}.2}\label{equationthmDeceiverDatasets.2}\\
					c_\Omega + c'
			%		<
			%		m - n
					-
					2 \, \mathbf{K}\left( \mathrm{P} \right) 
			%		- 
			%		\mathbf{K}\left( \mathrm{D}_a \right) 
			%		- 
			%		\mathbf{K}\left(  \delta \right)  
					- 
					\mathbf{O}\left( 1 \right)
					\leq \tag{\ref{equationthmDeceiverDatasets}.3}\label{equationthmDeceiverDatasets.3}\\
					m
					-
					2\, \mathbf{K}\left( \mathrm{P} \right) 
					- 
					n
					- \mathbf{O}\left( \log\left( n \right) \right)
					%		- 
					%		\mathbf{K}\left(  \delta \right)  
					- 
					\mathbf{O}\left( 1 \right)
					\leq \tag{\ref{equationthmDeceiverDatasets}.4}\label{equationthmDeceiverDatasets.4}\\
					m
					-
					\mathbf{K}\left( \mathrm{P} \right) 
					- 
					\mathbf{K}\left( \mathrm{D}_a \right) 
					%		- 
					%		\mathbf{K}\left(  \delta \right)  
					- 
					\mathbf{O}\left( 1 \right)
					< \tag{\ref{equationthmDeceiverDatasets}.5}\label{equationthmDeceiverDatasets.5}\\
					\mathbf{K}\left(  \mathrm{M}_{ \left( \mathrm{P} , \mathrm{D}_{ total } \right) }  \right) \nonumber
%					\tag{\ref{equationthmDeceiverDatasets}.6}
%					\label{equationthmDeceiverDatasets.6}
					\text{ ,}
				\end{align}
			\end{minipage}\label{equationthmDeceiverDatasets}
		\end{equation}
		\noindent where:
		\begin{itemize}
			\item Step~\ref{equationthmDeceiverDatasets.1} follows from Equations~\ref{equationlemmaD_a2} and~\ref{equationlemmaD_a3} in Lemma~\ref{lemmaD_a};
			\item Steps~\ref{equationthmDeceiverDatasets.2} and~\ref{equationthmDeceiverDatasets.3} follow from Equation~\ref{equationInequalityforc};
			\item Step~\ref{equationthmDeceiverDatasets.4} follows from Equation~\ref{equationlemmaD_a2} in Lemma~\ref{lemmaD_a};
			\item Step~\ref{equationthmDeceiverDatasets.5} follows from Equation~\ref{equationlemmaD_total} in Lemma~\ref{lemmaD_total}.
		\end{itemize}
		Hence, from basic properties of $ \mathbf{K}\left( \cdot \right) $ and Equation~\ref{equationthmDeceiverDatasets}, we know $ m $ can be chosen sufficiently greater than $ n $ so that there is $ C' > 0 $ such that 
		\begin{equation}
			\begin{aligned}
				\mathbf{K}\left( \mathrm{M}_{ \left( \mathrm{P} , \mathrm{D}_{ total } \right) } \, \middle\vert \, \left< \mathrm{D}_a , \mathrm{P} , \mathrm{M}_{ \left( \mathrm{P} , \mathrm{D}_a \right) }  \right> \right) 
				\geq \\
				\mathbf{K}\left( \mathrm{M}_{ \left( \mathrm{P} , \mathrm{D}_{ total } \right) } \right)
				-
				\mathbf{K}\left(  \mathrm{D}_a , \mathrm{P} , \mathrm{M}_{ \left( \mathrm{P} , \mathrm{D}_a \right) }  \right)
				- \mathbf{O}(1)
				\geq \\
				\mathbf{K}\left( \mathrm{M}_{ \left( \mathrm{P} , \mathrm{D}_{ total } \right) } \right)
				-
				\big(
				\mathbf{K}\left( \mathrm{P} \right) 
				+ \mathbf{K}\left( \mathrm{D}_a \right)
				+ \mathbf{K}\left(  \mathrm{M}_{ \left( \mathrm{P} , \mathrm{D}_a \right) }  \right)
				+ \mathbf{O}\left( 1 \right)
				\big)
				> \\
				C'
				\text{ .}
			\end{aligned}
		\end{equation}
		Therefore, since $ m $ can increase independently of $ n $ and $ \Omega \upharpoonright^m_n $ is incompressible with respect to
		$ \Omega \upharpoonright_n $, we have that Definition~\ref{defUnpredictablemodels} is satisfied.
		Additionally, together with the fact that $ \mathrm{D}_{ total } $ satisfies Lemma~\ref{lemmaD_total} by construction of program $ q $, we will have that Definition~\ref{defUnpredictabledeceiver} is satisfied.
%		Since $ c' $ was arbitrarily large and $ \mathrm{D}_a $ satisfies Lemma~\ref{lemmaD_a}, we can choose the minimal and sufficiently large value of $ c' $ so that 
%		$ \mathbf{K}\left( \mathrm{P} \right) 
%		%		- 
%		%		\mathbf{K}\left( \mathrm{D}_a \right) 
%		+ 
%		\mathbf{K}\left(  \delta \right) 
%		+
%		\mathbf{K}\left( \mathrm{M}_{ \left( \mathrm{P} , \mathrm{D}_a \right) } \right)
%		+ 
%		\mathbf{O}\left( 1 \right)
%		< 
%		\mathbf{K}\left(  \mathrm{M}_{ \left( \mathrm{P} , \mathrm{D}_{ total } \right) }  \right) $.
%		Furthermore, from the construction of the program $ h' $ in the proof of Lemma~\ref{label} with $ c'' = 0 $, we have it that 
%		$ \mathbf{K}\left( \mathrm{M}_{ \left( \mathrm{P} , \mathrm{D}_a \right) } \right) 
%		\leq
%		\mathbf{K}\left( \mathrm{P} \right)
%		+
%		\mathbf{K}\left( \mathrm{D}_{total} \right)
%		+
%		\mathbf{K}\left( \delta \right)
%		+
%		\mathbf{O}\left( 1 \right)
%		\leq
%		\mathbf{K}\left( \mathrm{M}_{ \left( \mathrm{P} , \mathrm{D}_{total} \right) } \right)
%		$.
%		From our construction of $ c $, since $ f\left( n \right) $  depends linearly on $ \mathbf{K}\left( \mathrm{P} \right) $ and $ \mathbf{K}\left( \delta \right) $, we have it that $ c $ depends linearly on $ \mathbf{K}\left( \mathrm{P} \right) $ and $ \mathbf{K}\left( \delta \right) $.
		Finally, 
		we will have that Equation~\ref{equationthmDeceiverDatasets3} holds because
		%		Lemma~\ref{label} and our construction of $ \mathrm{D}_{total} $, 
%		that 
		\begin{equation}
		\begin{minipage}{0.91\linewidth}
			\begin{align}
				\left| \mathrm{D}_a \right| 
				\geq \tag{\ref{equationfinalintheproofthm1}.1}\label{equationfinalintheproofthm1.1}\\
				BB\left( n \right) 
				%		\geq \\
				%		BB\left( m - c \right)
				\geq \tag{\ref{equationfinalintheproofthm1}.2}\label{equationfinalintheproofthm1.2}\\
				BB\Big(
				\mathbf{K}\left( \mathrm{D}_{total} \right) 
				- 
				\big(
				\mathbf{K}\left( \mathrm{P} \right)
				+ \mathbf{O}\left( \log\left( n \right) \right)
				+
				c
		%		+ \mathbf{O}\left( \log\left( c \right) \right)
				+
				\mathbf{O}\left( 1 \right)
				\big)
				\Big)
				\geq \tag{\ref{equationfinalintheproofthm1}.3}\label{equationfinalintheproofthm1.3}\\
				BB\Big(
				\mathbf{K}\left( \mathrm{D}_{total} \right) 
				- 
				\mathbf{O}\big( \mathbf{K}\left( \mathrm{P} \right)
				\big)
				\Big) \nonumber
%				\tag{\ref{equationfinalintheproofthm1}.4}\label{equationfinalintheproofthm1.4}
				\text{ .}
			\end{align}
		\end{minipage}\label{equationfinalintheproofthm1}
		\end{equation}
		holds, 
		where: 
		\begin{itemize}
			\item Step~\ref{equationfinalintheproofthm1.1} follows from Equation~\ref{equationlemmaD_a1} in Lemma~\ref{lemmaD_a};
			\item Step~\ref{equationfinalintheproofthm1.2} follows from Lemma~\ref{lemmaBB1} and from Equation~\ref{equationLaststepthm1};
			\item and Step~\ref{equationfinalintheproofthm1.3} follows from Lemma~\ref{lemmaBB1} and from Equations~\ref{equationthm1Basicassumptiononn} and~\ref{equationInequalityforc}.
		\end{itemize}
	\end{proof}
\end{theorem}

\subsection{Probability of very large deceiving datasets}\label{sectionProbabilityofbigdata}

So far, we have only shown the existence of unpredictable deceivers.
The next question would be how likely they are.
More precisely, in the context of randomly generated computably constructible datasets, one could ask after the probability of an unpredictable deceiver once the size of the datasets becomes very large with respect to the algorithmic complexity of the learning algorithm. 
We demonstrate in the following Theorem~\ref{thmDeceivingprobability} that, for sufficiently large datasets, the occurrence of an unpredictable deceiver is more likely (except for a constant that only depends on the learning algorithm) than the occurrence of any other particular dataset of interest.

\begin{theorem}\label{thmDeceivingprobability}
	Let $ \mathrm{P} $ be an arbitrary learning algorithm that satisfies Definition~\ref{defLearningalgorithm} such that $ \mathcal{X}^m $ is computable (or, equivalently without loss of generality, $ \mathcal{X}^m = \emptyset $).
	Let $ \mathcal{X}_{ \mathbf{U} } $ be a universally distributed data generating source as in Definition~\ref{defUniversaldatageneratingprocess}.
%	Let $ \mathrm{P} $ be an arbitrary learning algorithm.
	Let the available datasets $ \mathrm{D}_a $'s (from which $ \mathrm{P} $ calculates optimal models) be sufficiently large so that there is $ k \in \mathbb{N} $ such that 
	\begin{equation}
	\begin{aligned}
	\left| \mathrm{D}_a \right| 
	\geq  
	k 
	=
	BB\Big(
	\mathbf{K}\left( \mathrm{D}'_{total} \right) 
	- 
	\mathbf{O}\big( \mathbf{K}\left( \mathrm{P} \right)
	\big)
	\Big)
	\text{ ,}
	\end{aligned}\label{equationthmDeceivingprobability1}
	\end{equation}
	\noindent where $ \mathrm{D}'_{total} $ is any dataset that satisfies Theorem~\ref{thmDeceiverDatasets} for $ \mathrm{P} $.
	Then,  there are sufficiently large datasets $ \mathrm{D}_{ total } = \left( \mathrm{D}_a \, , \, \mathrm{D}_{new} \right) $ whose algorithmic probability (or the universal a priori probability) of the respective $ \mathrm{D}_a $ being an unpredictable deceiver (as in Definition~\ref{defUnpredictabledeceiver})
%	(which satisfy Theorem~\ref{thmDeceiverDatasets} with $ \left| \mathrm{D}_a \right| \geq k $) 
	is higher (except for a multiplicative constant that only depends on $ \mathbf{K}\left( \mathrm{P} \right) $) than the algorithmic probability (or the universal a priori probability) of any other particular dataset of size $ \geq k $,
	where the datasets $ \mathrm{D}_{ total } $ are generated by $ \mathcal{X}_{ \mathbf{U} } $ in the (computable) learning processes
	$ \mathcal{L}\left( \mathcal{X}_{ \mathbf{U} } , \mathcal{X}^m , f_{per} , \mathbf{F} , \mathrm{P} \right) $.
	
	\begin{proof}
		Let $ \mathrm{D}'_{total} = \left( \mathrm{D}'_a , \mathrm{D'_{new}} \right) $ be a particular dataset that satisfies Theorem~\ref{thmDeceiverDatasets} for the learning algorithm $ \mathrm{P} $.
		Therefore, we have that the size of $ \mathrm{D}'_a $ satisfies Equation~\ref{equationthmDeceivingprobability1}.
%		such that $ \mathbf{K}\left( \left| \mathrm{D}_a \right| \right) \geq \mathbf{K}\left( \mathrm{D}'_{total} \right) $.
%		Therefore, we know it also satisfies Lemma~\ref{label} with constants $ c'' $ and $ c''' $ such that $ \mathbf{K}\left(  \mathrm{D}'_{ total }  \right)
%		\leq
%		\log_2\left(  \left| \mathrm{D}'_{ a } \right| \right) $, where $ \mathrm{D}'_a $ is the available dataset in $ \mathrm{D}'_{total} $.
		Now, respectively from basic inequalities in AIT and Lemma~\ref{lemmaBB1}, 
		we know that for every $ \mathrm{D}_a $ with $ \left| \mathrm{D}_a \right| \geq k $, if
		$  k  
		= 
		BB\Big(
		\mathbf{K}\left( \mathrm{D}'_{total} \right) 
		- 
		\mathbf{O}\big( \mathbf{K}\left( \mathrm{P} \right)
		\big)
		\Big) $, then
		\begin{equation}\label{equationbasicthm2.1}
			\begin{aligned}
				\mathbf{K}\left( \mathrm{D}_{total} \right) 
		%		+ \mathbf{K}\left( \left| \mathrm{D}_a \right| \right) 
				+ \mathbf{O}\left( 1 \right)
				\geq \\
				\mathbf{K}\left( \left| \mathrm{D}_{total} \right| \right)
				+ \mathbf{O}\left( 1 \right) 
				\geq \\
				\mathbf{K}\left( \left| \mathrm{D}_a \right| \right)
				> \\
				\mathbf{K}\left( \mathrm{D}'_{total} \right) 
				- 
				\mathbf{O}\big( \mathbf{K}\left( \mathrm{P} \right)
				\big)
			\end{aligned}
		\end{equation}
		and 
		\begin{equation}
%		\begin{minipage}{0.9\linewidth}
				\begin{aligned}
				\mathbf{K}\left( \mathrm{D}_a \right) + \mathbf{O}\left( 1 \right)
				\geq \\ %\tag{\ref{equation2inthm2}.1}\\
				\mathbf{K}\left( \left| \mathrm{D}_a \right| \right) 
				\geq \\ %\tag{\ref{equation2inthm2}.2}\\
				\mathbf{K}\left( \mathrm{D}'_{total} \right) 
				- 
				\mathbf{O}\big( \mathbf{K}\left( \mathrm{P} \right)
				\big) %\tag{\ref{equation2inthm2}.3}
				 \text{ ,}
			\end{aligned}\label{equationbasicthm2.2}
%		\end{minipage}\label{equation2inthm2}
		\end{equation} 
		 where $ \mathrm{D}_{total} = \left( \mathrm{D}_a \, , \, \mathrm{D}_{new} \right) $ is any arbitrary dataset that extends the available dataset $ \mathrm{D}_a $.
		Hence, from our assumption on the size of the available datasets (in particular, Equation~\ref{equationthmDeceivingprobability1}) we will have that
		for any $ \mathrm{D}_{total} = \left( \mathrm{D}_a \, , \, \mathrm{D}_{new} \right) $,
		\begin{equation}\label{equationbasic2thm2.1}
			\frac{ 1 }{ 2^{ \left(  
					\mathbf{K}\left( \mathrm{D}_{total} \right) 
					%		+ \mathbf{K}\left( \left| \mathrm{D}_a \right| \right) 
					+ \mathbf{O}\left( 1 \right) 
					\right) } }
			\leq
			\frac{ 1 }{ 2^{ \left(  
					\mathbf{K}\left( \mathrm{D}'_{total} \right) 
					- 
					\mathbf{O}\big( \mathbf{K}\left( \mathrm{P} \right)
					\big) 
					\right) } }
		\end{equation} 
		holds because of Equation~\ref{equationbasicthm2.1}
		and
		\begin{equation}\label{equationbasic2thm2.2}
		\frac{ 1 }{ 2^{ \left(  
				\mathbf{K}\left( \mathrm{D}_a \right) + \mathbf{O}\left( 1 \right) 
				\right) } }
		\leq
		\frac{ 1 }{ 2^{ \left(  
				\mathbf{K}\left( \mathrm{D}'_{total} \right) 
				- 
				\mathbf{O}\big( \mathbf{K}\left( \mathrm{P} \right)
				\big) 
				\right) } }
		\end{equation} 
		holds because of Equation~\ref{equationbasicthm2.2}.
%		Let $ \delta $ be any arbitrarily fixed value that only depends on $ \mathrm{P} $ and it is a value that satisfies Theorem~\ref{thmDeceiverDatasets} for $ \mathrm{P} $.
		Note that the proof of Theorem~\ref{thmDeceiverDatasets} guarantees that there is at least some $ p' $ such that $ \mathbf{U}\left( p' \right) = \mathrm{D}'_{total} $.
		Therefore, from Definition~\ref{defUniversaldatageneratingprocess} and from the algorithmic coding theorem as in Equation~\ref{equationACT}, we have it that 
		\begin{equation}\label{equationACTinthm2}
			\mathbf{P} \left[ \text{ ``dataset } \mathrm{D}'_{total} \text{ occur''} \right] \geq
			C \sum\limits_{ \mathbf{U}\left( p \right) 
				= \mathrm{D}'_{total} } \frac{ 1 }{ 2^{ \left| p \right| } }
			\geq
			\frac{ 1 }{ 2^{ \left(  
					\mathbf{K}\left( \mathrm{D}'_{total} \right) 
					\right) } }
			\text{ ,}
		\end{equation} 
		where $ C $ is an independent constant.
		Finally, from Equations~\ref{equationbasic2thm2.1},~\ref{equationbasic2thm2.2}, and the algorithmic coding theorem, we have it that 
%		 $ 
%		\[
%		 2^{ - \left(  
%		 	\mathbf{K}\left( \mathrm{D}'_{total} \right) 
%		 	- 
%		 	\mathbf{O}\big( \mathbf{K}\left( \mathrm{P} \right)
%		 	\big) 
%	 	\right) }  
%	 	\]
	 	\begin{equation}
		 	\frac{ 1 }{ 2^{ \left(  
		 			\mathbf{K}\left( \mathrm{D}'_{total} \right) 
		 			- 
		 			\mathbf{O}\big( \mathbf{K}\left( \mathrm{P} \right)
		 			\big) 
		 			\right) } }
	 	\end{equation}
%	 	$\todo{fix} 
	 	is an upper bound for the algorithmic probability (or the universal a priori probability) of occurrence of any dataset (either $ \mathrm{D}_{total} $ or $ \mathrm{D}_a $) of size greater than or equal to $ k $ that is generated by $ \mathcal{X}_{ \mathbf{U} } $,
	 	where $ \mathrm{D}'_a $ is already known to be an unpredictable deceiver because by assumption $ \mathrm{D}'_{total} = \left( \mathrm{D}'_a , \mathrm{D'_{new}} \right) $ satisfies Theorem~\ref{thmDeceiverDatasets}.
	\end{proof}
	
\end{theorem}

%In the case of computable learning processes, we can improve Theorem~\ref{label} and demonstrate in Corollary~\ref{label} that the occurrence of deceiving datasets is as likely to occur as the learning algorithm itself.
%
%\begin{corollary}
%	Let $ \mathrm{P} $ be an arbitrary learning algorithm.
%	Let $ k $ be the size of the available datasets $ \mathrm{D}_a $'s (which $ \mathrm{P} $ may have access to calculate optimal models) such that 
%	$ \mathbf{K}\left( \left| \mathrm{D}_a \right| \right) \geq \mathbf{K}\left( \mathrm{D}_{total} \right) $, where $ \mathrm{D}_{total} $ is any dataset that satisfies Theorem~\ref{label}.
%	Then,  there are sufficiently large datasets $ \mathrm{D}_{total} $'s whose algorithmic probability of being deceivers that satisfy Theorem~\ref{label} is of the same order of the algorithmic probability of the learning algorithm, if the learning process is computable (i.e., free of stochasticity).
%\end{corollary}

\section{Discussion}\label{sectionDiscussion}

%\section{Deceiving algorithms into simple optimal solutions}\label{key}

%Theorem~\ref{thmDeceiverDatasets} demonstrates the existence of unpredictable deceivers for a given learning algorithm.
%Furthermore, as long as the datasets are randomly generated by computable processes, Theorem~\ref{thmDeceivingprobability} shows that if a particular dataset with a sufficiently large size is under analysis, then the probability that it is an unpredictable deceiver is larger than the probability of 

We have investigated the algorithmic probability of sufficiently large datasets that are unpredictable deceivers for a given learning algorithm.
Estimating the optimal model for these datasets does not guarantee that the  model will continue to be globally optimal.
For such kind of inverse problem \cite{Zenil2020cnat} (i.e., the problem of finding the unknown cause of a known effect, or finding the unknown generative model of the underlying process that generated the known phenomena), inferring the best solution is analogous to inferring the optimal model.
Our results assure that computable learning processes are not, in general, capable of solving inverse problems. This is also the case when the algorithm accounts for a bias-toward-simplicity given by the algorithmic coding theorem. 
%---which may render our results to appear counter-intuitive at first glance.
We have shown that, given a bias toward low algorithmic complexity (or high algorithmic probability), \emph{no} computable learning process can in general give reliable predictions for any randomly generated computably constructible collection of data. That is, there is always an optimization limit such as with the no free lunch theorems discussed in Section~\ref{sectionMachinelearninginlargedatasets}.

The condition in Equation~\ref{equationthmDeceivingprobability1} imposing a lower bound on the size of the datasets guarantees that the algorithmic complexity of any other set containing fresh data will be sufficiently high so that it can render the original dataset a deceiver that induces the learning algorithm to find a locally optimal model whose algorithmic complexity is relatively much smaller. 
One can understand those deceivers for inverse problems that (as shown in Theorems~\ref{thmDeceiverDatasets} and~\ref{thmDeceivingprobability}) arise from a wide enough gap between the algorithmic complexity of the underlying generative model of the phenomena and the algorithmic complexity of the problem solver (in our case, the learning algorithm) as \textit{the cause} of the simplicity bubble effect. That is, an underlying data generating process (mechanism or external source) traps an algorithm (or a formal theory) into a \emph{simplicity bubble} if the data made available by the data generating process is complex enough to ensure that any optimal model proposed by the algorithm (or formal theory) is not complex enough to successfully approximate the globally optimal model that actually corresponds to any fresh data that the algorithm may receive.

%\begin{SBE}
%	We \emph{informally} say that an underlying data generating process (mechanism or external source) is trapping an algorithm (or a formal theory) into a \emph{simplicity bubble} if the data made available by this data generating process is complex enough to ensure that any optimal model proposed by the algorithm (or formal theory) is not complex enough to successfully approximate the globally optimal model that actually corresponds to any fresh data that the algorithm may receive.
%\end{SBE}

As demonstrated by Theorem~\ref{thmDeceivingprobability}, we can formalise the simplicity bubble effect for the case of the computable learning processes $ \mathcal{L}\left( \mathcal{X}_{ \mathbf{U} } , \mathcal{X}^m , f_{per} , \mathbf{F} , \mathrm{P} \right) $ that we have studied in this article:

\begin{definition}[\textbf{Simplicity Bubble Effect} for universally distributed data generating processes]\label{defSBEforuniversallydistributeddatageneratingsources}
	Let $ \mathcal{L}\left( \mathcal{X}_{ \mathbf{U} } , \mathcal{X}^m , f_{per} , \mathbf{F} , \mathrm{P} \right) $ be a computable learning process as in Definition~\ref{defComputablelearningprocess}, where $ \mathcal{X}_{ \mathbf{U} } $ is a universally distributed data generating source as in Definition~\ref{defUniversaldatageneratingprocess}.
	We say that the data generating source $ \mathcal{X}_{ \mathbf{U} } $ is trapping the learning algorithm $ \mathrm{P} $ into a \emph{simplicity bubble} if the algorithmic complexity $ \mathbf{K}\left( \mathrm{D}_a \right) $ of the available dataset $ \mathrm{D}_a $ is sufficiently large so that any optimal model $ \mathrm{M}_{ \left( \mathrm{P} , \mathrm{D}_a \right) } $  (as in Definition~\ref{defOptimalmodelfromavailabledata} and Notation~\ref{defModelcalculatedbylearningalgorithm}) is \emph{not} a globally optimal model (as in Definition~\ref{defGlobaloptima}) and it has \emph{not} enough necessary algorithmic information to predict a globally optimal model $ \mathrm{M}_{ \left( \mathrm{P} , \mathrm{D}_{total} \right) } $.
\end{definition}

While we prove that the simplicity bubble effect will always occur for sufficiently large datasets in Theorems~\ref{thmDeceiverDatasets} and \ref{thmDeceivingprobability}, it is an open question that warrants more research whether or not deceivers can arise from much smaller sets whose underlying generative model indeed has a sufficiently large algorithmic complexity in comparison to that of the learning algorithm.
If the answer is positive, this would reveal instances of the simplicity bubble effect that occur not only in very large datasets, but also in a much wider range of inverse problems.

Future research is also warranted to investigate the simplicity bubble effect for types of learning processes other than $ \mathcal{L}\left( \mathcal{X}_{ \mathbf{U} } , \mathcal{X}^m , f_{per} , \mathbf{F} , \mathrm{P} \right) $.
It is possible that deceiving phenomenon also occurs in other types of stochastic data generating sources
% $ \mathcal{X}_{ \mathrm{D} } $, in types of 
such as models other than ones for regression tasks.
In particular, this an important subject to be tackled in the context of data generating processes biased towards simplicity constrained to the space of resource-bounded Turing machines as opposed to the resource-unbounded case in the traditional algorithmic coding theorem that we studied in this article.

\subsection{Avoiding the simplicity bubble effect}\label{sectionAvoidingdeceivers}

Another immediate question is whether or not there are conditions that would avoid deceivers.
A trivial example in which a simplicity bubble is not expected to occur (with probability as high as one wishes) is for independent and identically distributed (i.i.d.) stochastic data generating sources.
For example, let the datasets $ \mathrm{D} $ be composed of data points of the form of the pair $ \left( X  , \phi_{ \mathrm{D} }\left( X \right) \right) $, where:
$ \mathcal{X}'_{ \mathrm{D} } $ is a stochastic data generating source of the datasets defined upon another stochastic source $ \mathcal{X} $; 
$ X $ is a random variable of the i.i.d. uniform stochastic source $ \mathcal{X} $ that generates values in $ \left\{ 0 , 1 \right\} $; and
$ \phi_{ \mathrm{D} }\left( X \right) \in Y $ is the empirical probability distribution of the value randomly generated by $ \mathcal{X} $ in the dataset $ \mathrm{D} $ such that
\begin{equation}
\begin{array}{lccc}
\mathrm{M} \colon & X_{ \mathrm{D} } \subseteq \left\{ 0 , 1 \right\} & \to & Y \subseteq \left( \mathbb{Q} \cap \left[ 0 , 1 \right] \right) \\
&  x & \mapsto & \mathrm{M}\left( x \right) = \theta\left( x \right)
\end{array}
\end{equation} 
is a model for a regression task that tries to predict the empirical probability distribution of each value $ X $ in the datasets $ \mathrm{D} $;
and $ \theta\left( x \right) $ is a parameter value in $ \left[ 0 , 1 \right] \subset \mathbb{R} $ such that $ \boldsymbol{ \theta } = \left\{ \theta\left( 0 \right) , \theta\left( 1 \right) \right\} $ is the set of parameters of the model $ \mathrm{M} $.
Let the performance measure be defined by the KL divergence and the optimality criterion by any arbitrarily small $ \epsilon $ below which the KL divergence between the probability distribution $ \boldsymbol{ \theta } $ and the empirical probability distribution $ \left( \phi_{ \mathrm{D} }\left( 0 \right) , \phi_{ \mathrm{D} }\left( 1 \right) \right) $ should be in order to $ \boldsymbol{ \theta } $ be considered optimal.
(See also Sections~\ref{sectionMachinelearning} and~\ref{sectionComputabilitytheoryandmachinelearning}).

In this example, it is well known that one can find a learning algorithm that can approximate the globally optimal model $ \boldsymbol{ \theta' } = \left\{ 1/2 , 1/2 \right\} $ as close as one wishes as the size of the randomly generated datasets increases.
This trivially holds because of the law of large numbers \cite{Cover2005}.
In this example, we know beforehand that the simplicity bubble effect will never occur and that deceivers (as in Definition~\ref{defDeceivingdataset}) will never occur in the long run.
In order to check why our present results do not contradict this example, just note that $ \mathcal{X}'_{ \mathrm{D} } $ does \emph{not} satisfies Definition~\ref{defUniversaldatageneratingprocess} for which Theorem~\ref{thmDeceivingprobability} is proven to hold.
This is because for every deceiving dataset $ \mathrm{D}_a $ that satisfies Definition~\ref{defDeceivingdataset}, we will have from the law of large numbers that
\begin{equation}
\lim\limits_{ \left| \mathrm{D}_a \right| \to \infty }
\mathbf{P} \left[ \text{ ``dataset } \left( \mathrm{D}_a , \mathrm{D}_{new} \right) \text{ occur''} \right] = 
0
%\text{ ,}
\end{equation}
holds such that this convergence is exponential with the value of $ \epsilon $ and the size $ \left| \mathrm{D}_{new} \right| $ of the fresh dataset \cite{Cover2005}.
This directly contradicts Equation~\ref{equationDefinitionUniversalprobability}, in which the convergence cannot occur faster than the algorithmic probability of the shortest program that outputs $ \mathrm{D}_{new} $ from $ \mathrm{D}_a $.
Hence, $ \mathcal{X}'_{ \mathrm{D} } $ is not a universally distributed data generating source.

In addition, the fact that $ \mathcal{X} $ is i.i.d. by definition guarantees that it is stationary, and therefore there is an upper bound for the algorithmic complexity $ \mathbf{K}\left( \boldsymbol{ \theta' } \right) $ of the global optimal probability distribution $ \boldsymbol{ \theta' } $.
This motivates the investigation of additional conditions (such as this upper bound) that would avoid deceivers even in the case the data generating source is indeed universally distributed (which was not the case of the above example for $ \mathcal{X}'_{ \mathrm{D} } $).

In addition to keeping the assumption of a universal bias toward simplicity as in Definition~\ref{defUniversaldatageneratingprocess}, one would need another assumption that avoids deceivers like those demonstrated in Section~\ref{sectionResults}.
To this end, note that the first key step in Lemma~\ref{lemmaD_a}, fundamental for the proof of Theorem~\ref{thmDeceiverDatasets}, relies on $ n $ being arbitrarily large. This, in turn, leads $ m $ in Lemma~\ref{lemmaD_total} to also be larger. 
These two values increase the algorithmic complexity of the deceiving dataset.
We will propose (in the next Section~\ref{sectionComplexitycaging}) an immediate way to avoid this by assuming a condition that ensures that the algorithmic complexity of the datasets cannot be much greater than that of the learning algorithm.
Furthermore, the algorithmic complexity of the optimal model calculated from the available dataset $ \mathrm{D}_a $ must be sufficiently smaller than the algorithmic complexity of $ \mathrm{D}_a $ so as to avoid overfitting \cite{HernandezOrozco2021}.

\subsection{A new hypothesis in order to avoid the simplicity bubble}\label{sectionComplexitycaging}

When solving inverse problems from empirical data, as discussed in the end of Section~\ref{sectionLearningprocesses}, the satisfiability of the necessary conditions (e.g., the bias toward simplicity assured by Definition~\ref{defUniversaldatageneratingprocess}) are always of both empirical and mathematical nature.
We now introduce a new assumption on the underlying generative models of the datasets with the purpose of guaranteeing that the chosen computable method can indeed find the global optimal solution for a given available dataset. Essentially, the complexity of the phenomena about which the learning algorithms can make predictions cannot be greater than (up to a small constant) the complexity of the learning algorithms themselves. To acheive this, we can try to ensure that the complexity of the optimal models (which are produced by the learning algorithms from the available datasets) are smaller than the complexity of the datasets.
In other words, given that the data generating processes are biased toward low complexity, we can impose limits that ``cage'' the complexity of the data on which predictions can be made through two methods. The first condition is to ensure that the data used for prediction is limited by the complexity of the learning algorithm. The second condition is to ensure that the complexity of the optimal models given by the learning algorithm remains below the complexity of any possible available data.
Then, predictions would only be reliable in a computably generated universe if the complexity of the phenomena is trapped into this ``complexity cage''. 

\begin{CCP}
	Let $ \mathrm{P} $ be an arbitrary learning algorithm that satisfies Definition~\ref{defLearningalgorithm} such that $ \mathcal{X}^m $ is computable (or, equivalently without loss of generality, $ \mathcal{X}^m = \emptyset $).
	Let $ \mathcal{X}_{ \mathbf{U} } $ be a universally distributed data generating source as in Definition~\ref{defUniversaldatageneratingprocess}.
	We say $ \mathrm{P} $ can give a \emph{reliable solution} to inverse problems on a (randomly generated or deterministically) computably constructible dataset $ \mathrm{D}_{ total } = \mathrm{D}_a \cup \mathrm{D}_{new} $ 
	%	where $ \mathrm{D}_a $ is an available dataset from which $ \mathrm{P} $ outputs the predicting optimal models $ \mathrm{M}_{ \left( \mathrm{P} , \mathrm{D}_a \right) } $ and $ \mathrm{D}_{new} $ is the dataset composed of the fresh data that $ \mathrm{P} $ has no access to in the learning process, 
	iff there is a constant $ c $ that depends only on $ \mathrm{P} $ such that for every $ \mathrm{D}_{total} $:
	\begin{enumerate}
		\item $  \mathbf{K}\left( \mathrm{D}_{total} \right) \leq \mathbf{K}\left( \mathrm{P} \right) + c $;

		\item or $  \mathbf{K}\left( \mathrm{M}_{ \left( \mathrm{P} , \mathrm{D}_{total} \right) } \right) \leq \mathbf{K}\left( \mathrm{P} \right) + c $, where
		$ \mathrm{M}_{ \left( \mathrm{P} , \mathrm{D}_{total} \right) } $ is an arbitrary globally optimal model that satisfies Definition~\ref{defGlobaloptima}; 
		
		\item or both hold.

	\end{enumerate}
\end{CCP}

While the above ``complexity caging'' strategy clearly avoids the consequences implied by the proofs presented in this article, future research is necessary to demonstrate whether or not it is capable of avoiding every possible deceiving dataset.

It remains an open problem whether or not there are  mathematical conditions that, if satisfied, guarantee that the prediction on empirical data is reliable. As such conditions would be based on algorithmic information dynamics, this suggests that stronger versions of machine learning based on computability theory and algorithmic information theory may overcome the limitations of statistical machine learning (including deep learning).
This becomes evinced in the context of third-party agents or systems that may try to steer or perturb real-world data in order to intentionally mislead machine learning algorithms.
In this case, to achieve reliable predictions using the complexity caging principle, the datasets would have to be analyzed to ensure that its algorithmic complexity is lower than that of the learning algorithm itself. 
This could be done with novel pre-processing methods (advancing on those discussed in Section~\ref{sectionMachinelearninginlargedatasets}), which are based on approximations to the value of algorithmic complexity that satisfy the complexity caging principle. 
Such a method has the potential to complement and extend generalist (or universalist) machine learning methods to ensure that predictions are not only reliable for the available data, but also for new yet unavailable phenomena.

\section*{Funding}

Felipe S. Abrah\~{a}o acknowledges the partial support from São Paulo Research Foundation (FAPESP), grant \# $2021$/$14501$-$8$.
Authors acknowledge the partial support from: MCTI and CNPq through their individual grants to F. S. Abrah\~{a}o (302185/2021-6), K. Wehmuth (302211/2021-7), and F. Porto (PQ-313961/2018-2).



\bibliographystyle{elsarticle-num.bst}
\bibliography{3-CompleteRefs-Felipe.bib}

\begin{thebibliography}{10}
\expandafter\ifx\csname url\endcsname\relax
  \def\url#1{\texttt{#1}}\fi
\expandafter\ifx\csname urlprefix\endcsname\relax\def\urlprefix{URL }\fi
\expandafter\ifx\csname href\endcsname\relax
  \def\href#1#2{#2} \def\path#1{#1}\fi

\bibitem{Calude2017}
C.~S. Calude, G.~Longo,
  \href{http://link.springer.com/10.1007/s10699-016-9489-4}{The {Deluge} of
  {Spurious} {Correlations} in {Big} {Data}}, Foundations of Science 22~(3)
  (2017) 595--612.
\newblock \href {https://doi.org/10.1007/s10699-016-9489-4}
  {\path{doi:10.1007/s10699-016-9489-4}}.
\newline\urlprefix\url{http://link.springer.com/10.1007/s10699-016-9489-4}

\bibitem{Smith2020}
G.~Smith, \href{http://link.springer.com/10.1007/s42452-020-2862-5}{The paradox
  of big data}, SN Applied Sciences 2~(6) (2020) 1041.
\newblock \href {https://doi.org/10.1007/s42452-020-2862-5}
  {\path{doi:10.1007/s42452-020-2862-5}}.
\newline\urlprefix\url{http://link.springer.com/10.1007/s42452-020-2862-5}

\bibitem{Goodfellow2016}
I.~Goodfellow, Y.~Bengio, A.~Courville, Deep Learning, MIT Press, 2016,
  \url{http://www.deeplearningbook.org}.

\bibitem{Witten2017}
I.~H. Witten, E.~Frank, M.~A. Hall, C.~J. Pal (Eds.),
  \href{https://www.sciencedirect.com/science/article/pii/B9780128042915000052}{Data
  Mining}, fourth edition Edition, Morgan Kaufmann, 2017.
\newblock \href {https://doi.org/10.1016/C2015-0-02071-8}
  {\path{doi:10.1016/C2015-0-02071-8}}.
\newline\urlprefix\url{https://www.sciencedirect.com/science/article/pii/B9780128042915000052}

\bibitem{HernandezOrozco2021}
S.~Hernández-Orozco, H.~Zenil, J.~Riedel, A.~Uccello, N.~A. Kiani, J.~Tegnér,
  \href{https://www.frontiersin.org/articles/10.3389/frai.2020.567356/full}{Algorithmic
  {Probability}-{Guided} {Machine} {Learning} on {Non}-{Differentiable}
  {Spaces}}, Frontiers in Artificial Intelligence 3 (2021) 567356.
\newblock \href {https://doi.org/10.3389/frai.2020.567356}
  {\path{doi:10.3389/frai.2020.567356}}.
\newline\urlprefix\url{https://www.frontiersin.org/articles/10.3389/frai.2020.567356/full}

\bibitem{Calude2002}
C.~S. Calude, {Information and Randomness: An algorithmic perspective}, 2nd
  Edition, Springer-Verlag, 2002.

\bibitem{Bishop2006}
C.~M. Bishop, Pattern recognition and machine learning, Information science and
  statistics, Springer, New York, 2006.

\bibitem{Murphy2012}
K.~P. Murphy, Machine learning: a probabilistic perspective, 2012.

\bibitem{Nielsen2015}
M.~A. Nielsen, \href{http://neuralnetworksanddeeplearning.com}{Neural networks
  and deep learning}, Determination press, 2015.
\newline\urlprefix\url{http://neuralnetworksanddeeplearning.com}

\bibitem{Mitchell1997}
T.~M. Mitchell, Machine {Learning}, {McGraw}-{Hill} series in computer science,
  McGraw-Hill, New York, 1997.

\bibitem{Shcherbakov2013}
M.~Shcherbakov, A.~Brebels, N.~Shcherbakova, A.~Tyukov, T.~Janovsky, V.~Kamaev,
  A survey of forecast error measures, World Applied Sciences Journal 24 (2013)
  171--176.
\newblock \href {https://doi.org/10.5829/idosi.wasj.2013.24.itmies.80032}
  {\path{doi:10.5829/idosi.wasj.2013.24.itmies.80032}}.

\bibitem{Baharan2021}
B.~Mirzasoleiman, J.~A. Bilmes, J.~Leskovec,
  \href{http://proceedings.mlr.press/v119/mirzasoleiman20a.html}{Coresets for
  data-efficient training of machine learning models}, in: Proceedings of the
  37th International Conference on Machine Learning, {ICML} 2020, 13-18 July
  2020, Virtual Event, Vol. 119 of Proceedings of Machine Learning Research,
  {PMLR}, 2020, pp. 6950--6960.
\newline\urlprefix\url{http://proceedings.mlr.press/v119/mirzasoleiman20a.html}

\bibitem{Pereira2021}
R.~Pereira, Y.~Souto, A.~Chaves, R.~Zorilla, B.~Tsan, F.~Rusu, E.~Ogasawara,
  A.~Ziviani, F.~Porto,
  \href{https://doi.org/10.1145/3468791.3468806}{Djensemble: A cost-based
  selection and allocation of a disjoint ensemble of spatio-temporal models},
  Association for Computing Machinery, New York, NY, USA, 2021, p. 226–231.
\newline\urlprefix\url{https://doi.org/10.1145/3468791.3468806}

\bibitem{Wolpert1997}
D.~Wolpert, W.~Macready, No free lunch theorems for optimization, IEEE
  Transactions on Evolutionary Computation 1~(1) (1997) 67--82.
\newblock \href {https://doi.org/10.1109/4235.585893}
  {\path{doi:10.1109/4235.585893}}.

\bibitem{Scholkopf2021}
B.~Scholkopf, F.~Locatello, S.~Bauer, N.~R. Ke, N.~Kalchbrenner, A.~Goyal,
  Y.~Bengio, \href{https://ieeexplore.ieee.org/document/9363924/}{Toward
  {Causal} {Representation} {Learning}}, Proceedings of the IEEE 109~(5) (2021)
  612--634.
\newblock \href {https://doi.org/10.1109/JPROC.2021.3058954}
  {\path{doi:10.1109/JPROC.2021.3058954}}.
\newline\urlprefix\url{https://ieeexplore.ieee.org/document/9363924/}

\bibitem{Chaitin2004}
G.~Chaitin, {Algorithmic Information Theory}, 3rd Edition, Cambridge University
  Press, 2004.

\bibitem{Zenil2018}
H.~Zenil, S.~Hern{\'{a}}ndez-Orozco, N.~Kiani, F.~Soler-Toscano,
  A.~Rueda-Toicen, J.~Tegn{\'{e}}r,
  \href{http://www.mdpi.com/1099-4300/20/8/605}{{A Decomposition Method for
  Global Evaluation of Shannon Entropy and Local Estimations of Algorithmic
  Complexity}}, Entropy 20~(8) (2018) 605.
\newblock \href {https://doi.org/10.3390/e20080605}
  {\path{doi:10.3390/e20080605}}.
\newline\urlprefix\url{http://www.mdpi.com/1099-4300/20/8/605}

\bibitem{Zenil2019}
H.~Zenil, N.~A. Kiani, A.~A. Zea, J.~Tegn{\'{e}}r,
  \href{http://www.nature.com/articles/s42256-018-0005-0}{{Causal deconvolution
  by algorithmic generative models}}, Nature Machine Intelligence 1~(1) (2019)
  58--66.
\newblock \href {https://doi.org/10.1038/s42256-018-0005-0}
  {\path{doi:10.1038/s42256-018-0005-0}}.
\newline\urlprefix\url{http://www.nature.com/articles/s42256-018-0005-0}

\bibitem{Zenil2017a}
H.~Zenil, N.~A. Kiani, J.~Tegn{\'{e}}r,
  \href{http://dx.doi.org/10.1103/PhysRevE.96.012308}{{Low-algorithmic-complexity
  entropy-deceiving graphs}}, Physical Review E 96~(1) (2017) 012308.
\newblock \href {https://doi.org/10.1103/PhysRevE.96.012308}
  {\path{doi:10.1103/PhysRevE.96.012308}}.
\newline\urlprefix\url{http://dx.doi.org/10.1103/PhysRevE.96.012308}

\bibitem{Downey2010}
R.~G. Downey, D.~R. Hirschfeldt,
  \href{http://link.springer.com/10.1007/978-0-387-68441-3}{{Algorithmic
  Randomness and Complexity}}, Theory and Applications of Computability,
  Springer New York, New York, NY, 2010.
\newblock \href {http://arxiv.org/abs/http://doi.org/10.1007/978-0-387-68441-3}
  {\path{arXiv:http://doi.org/10.1007/978-0-387-68441-3}}, \href
  {https://doi.org/10.1007/978-0-387-68441-3}
  {\path{doi:10.1007/978-0-387-68441-3}}.
\newline\urlprefix\url{http://link.springer.com/10.1007/978-0-387-68441-3}

\bibitem{Li1997}
M.~Li, P.~Vit{\'{a}}nyi, {An Introduction to Kolmogorov Complexity and Its
  Applications}, 4th Edition, Texts in Computer Science, Springer, Cham, 2019.
\newblock \href {https://doi.org/10.1007/978-3-030-11298-1}
  {\path{doi:10.1007/978-3-030-11298-1}}.

\bibitem{Cover2005}
T.~M. Cover, J.~A. Thomas, {Elements of Information Theory}, John Wiley {\&}
  Sons, Inc., Hoboken, NJ, USA, 2005.
\newblock \href {http://arxiv.org/abs/ISBN 0-471-06259-6} {\path{arXiv:ISBN
  0-471-06259-6}}, \href {https://doi.org/10.1002/047174882X}
  {\path{doi:10.1002/047174882X}}.

\bibitem{Abrahao2017publishednat}
F.~S. Abrah{\~{a}}o, K.~Wehmuth, A.~Ziviani, {Algorithmic networks: Central
  time to trigger expected emergent open-endedness}, Theoretical Computer
  Science 785 (2019) 83--116.
\newblock \href {https://doi.org/10.1016/j.tcs.2019.03.008}
  {\path{doi:10.1016/j.tcs.2019.03.008}}.

\bibitem{Colbrook2022}
M.~J. Colbrook, V.~Antun, A.~C. Hansen,
  \href{https://pnas.org/doi/full/10.1073/pnas.2107151119}{The difficulty of
  computing stable and accurate neural networks: {On} the barriers of deep
  learning and {Smale}’s 18th problem}, Proceedings of the National Academy
  of Sciences 119~(12) (2022).
\newblock \href {https://doi.org/10.1073/pnas.2107151119}
  {\path{doi:10.1073/pnas.2107151119}}.
\newline\urlprefix\url{https://pnas.org/doi/full/10.1073/pnas.2107151119}

\bibitem{Chaitin2012}
G.~Chaitin,
  \href{http://www.worldscientific.com/doi/abs/10.1142/9789814374309{\_}0015}{{Life
  as Evolving Software}}, in: H.~Zenil (Ed.), A Computable Universe, World
  Scientific Publishing, Singapure, 2012, pp. 277--302.
\newblock \href {https://doi.org/10.1142/9789814374309_0015}
  {\path{doi:10.1142/9789814374309_0015}}.
\newline\urlprefix\url{http://www.worldscientific.com/doi/abs/10.1142/9789814374309{\_}0015}

\bibitem{Bennett1988}
C.~H. Bennett, Logical {Depth} and {Physical} {Complexity}, in: A
  {Half}-{Century} {Survey} on {The} {Universal} {Turing} {Machine}, Oxford
  University Press, Inc., USA, 1988, pp. 227--257.

\bibitem{Abrahao2016nat}
F.~S. Abrah{\~{a}}o, {The ``paradox'' of computability and a recursive relative
  version of the Busy Beaver function}, in: C.~Calude, M.~Burgin (Eds.),
  {Information and Complexity}, 1st Edition, World Scientific Publishing,
  Singapure, 2016, Ch.~1, pp. 3--15.
\newblock \href {https://doi.org/10.1142/9789813109032_0001}
  {\path{doi:10.1142/9789813109032_0001}}.

\bibitem{Chaitin1975}
G.~Chaitin, {A Theory of Program Size Formally Identical to Information
  Theory}, Journal of the ACM 22~(3) (1975) 329--340.
\newblock \href {https://doi.org/10.1145/321892.321894}
  {\path{doi:10.1145/321892.321894}}.

\bibitem{Zenil2020cnat}
H.~Zenil, N.~Kiani, F.~Abrah{\~{a}}o, J.~Tegn{\'{e}}r, {Algorithmic Information
  Dynamics}, Scholarpedia Journal 15~(7) (2020) 53143.
\newblock \href {https://doi.org/10.4249/scholarpedia.53143}
  {\path{doi:10.4249/scholarpedia.53143}}.

\end{thebibliography}


%\begin{thebibliography}{00}
%%% Text of bibliographic item
%
%\bibitem{}
%
%\end{thebibliography}
\end{document}